\documentclass[aps,prl,twocolumn,superscriptaddress,longbibliography]{revtex4}
\usepackage{graphicx}
\usepackage{amsmath}
\usepackage{amssymb}
\usepackage{amsfonts}
\usepackage{bbm,bm}
\usepackage{lineno}
\usepackage{setspace}

\usepackage{xcolor}

 \ifx\pdftexversion\undefined
\usepackage[dvips]{hyperref}
\else
\usepackage{hyperref}
\fi
\hypersetup{
  colorlinks = true, linkcolor = blue
}
\begin{document}
\preprint{}

\title{
Suppression of Bogoliubov momentum pairing and emergence of non-Gaussian correlations in ultracold interacting Bose gases}

\author{Jan-Philipp Bureik}
\email{These authors contributed equally.}
\affiliation{Universit\'e Paris-Saclay, Institut d'Optique Graduate School, CNRS, Laboratoire Charles Fabry, 91127, Palaiseau, France}
\author{Ga\'etan Herc\'e}
\email{These authors contributed equally.}
\affiliation{Universit\'e Paris-Saclay, Institut d'Optique Graduate School, CNRS, Laboratoire Charles Fabry, 91127, Palaiseau, France}
\author{Maxime Allemand}
\affiliation{Universit\'e Paris-Saclay, Institut d'Optique Graduate School, CNRS, Laboratoire Charles Fabry, 91127, Palaiseau, France}
\author{Antoine Tenart}
\affiliation{Universit\'e Paris-Saclay, Institut d'Optique Graduate School, CNRS, Laboratoire Charles Fabry, 91127, Palaiseau, France}
\author{Tommaso Roscilde}
\affiliation{Univ Lyon, Ens de Lyon, CNRS, Laboratoire de Physique, F-69342 Lyon, France}
\author{David Cl\'ement}
\affiliation{Universit\'e Paris-Saclay, Institut d'Optique Graduate School, CNRS, Laboratoire Charles Fabry, 91127, Palaiseau, France}

\begin{abstract}
Strongly correlated quantum matter \cite{Sachdevbook2023} -- such as interacting electron systems \cite{Morosan2012} or interacting quantum fluids  \cite{Adams2012, Bloch2022, Chevy2016} -- possesses  properties that cannot be understood in terms of linear fluctuations and free quasi-particles \cite{georges1996, metzner2012}. Quantum fluctuations in these systems are indeed large and generically exhibit non-Gaussian statistics \cite{Georges2013, Chang2014} -- a property captured only by inspecting high-order correlations, whose quantitative reconstruction poses a formidable challenge to both experiments and theory alike. A prime example of correlated quantum matter is the strongly interacting Bose fluid, realized by superfluid Helium \cite{Stringaribook, leggettbook} and, more recently, ultra-cold atoms \cite{Makotyn2014, Fletcher2016,eismann2016, Klauss2017, Eigen2018, Yan2020}.
Here, we experimentally study interacting Bose gases from the weakly to the strongly interacting regime through single-atom-resolved correlations in momentum space. We observe that the Bogoliubov pairing among modes of opposite momenta, emblematic of the weakly interacting regime \cite{Bogoliubov1947, tenart2021}, is suppressed as interactions become stronger. This departure from the predictions of Bogoliubov theory signals the onset of the strongly correlated regime, as confirmed by numerical simulations that highlight the role of non-linear quantum fluctuations in our system. 
Additionally, our measurements unveil a non-zero four-operator cumulant at even stronger interactions, which is a direct signature of non-Gaussian correlations. 
These results shed light on the emergence and physical origin of non-Gaussian correlations in ensembles of interacting bosons.
\end{abstract}

\maketitle

Quantum fluctuations induced by interactions play a prominent role in the equilibrium properties \cite{Sachdevbook2023,  Paschen2021} and/or in the dynamics \cite{Polkolnikov2011} of ensembles of interacting quantum particles. 
When fluctuations are weak, they can be captured by a linear theory which describes the quantum state of the system as a Gaussian state \cite{Gaussian_states}.  In Gaussian states any correlation can be expressed as a combination of the first two cumulants (mean and variance) of fluctuations. In contrast, strongly correlated quantum systems exhibit substantial quantum fluctuations that linear theories fail to account for \cite{georges1996, metzner2012, Chevy2016}. The unique nature of these systems reveals itself in (some) observables exhibiting non-Gaussian statistics, which are essential to understand the microscopic behavior \cite{Armijo2010, Schweigler2017, Rispoli2019} but are difficult to identify in both experiments and theories.

Bosons with pairwise contact interaction can realize a paradigmatic example of strongly correlated quantum systems, namely the strongly interacting Bose gas. In the regime of weak interactions, Bogoliubov theory \cite{Bogoliubov1947} -- leading to the celebrated Lee-Huang-Yang correction to the mean-field equation of state \cite{lee1957, navon2011, Jorgensen2018} -- treats interaction-induced quantum fluctuations \cite{pitaevskii1991} as linear fluctuations of a nearly perfect Bose-Einstein condensate (BEC). The fraction of atoms expelled from the BEC -- the quantum depletion \cite{lopes2017} -- is described in terms of two-mode-squeezed states for pairs of modes with opposite momenta -- a relevant example of a Gaussian bosonic state. This pairwise population of modes at opposite momenta (hereafter referred to as Bogoliubov momentum pairing) is the emblematic form of correlation of the weakly interacting regime, as recently confirmed experimentally \cite{tenart2021}.

When interactions become strong enough as to push the system away from the weakly interacting regime, the reconstruction of correlations going beyond those predicted by Bogoliubov theory becomes a challenge, for both experiments and theory. 
On the theory side, the main theoretical tool to treat strong correlations is offered by numerical calculations \cite{Ceperley1995,Svistunovbook} -- which nonetheless rarely address the correlations probing non-Gaussian statistics, because of their significant numerical cost. 
On the experimental front, ultracold atoms offer the possibility to control the strength of interactions, giving access to the rich showcase of quantum many-body phenomena exhibited by strongly interacting bosons \cite{fisher1989, bloch2008, Georgescu2014}. Resonant two-body interactions tuned with Feshbach resonances realize unitary Bose gases \cite{Chevy2016} hosting few-body Efimov physics \cite{Makotyn2014, Fletcher2016,eismann2016, Klauss2017, Eigen2018, Yan2020}, which are nonetheless plagued by large three-body losses \cite{Chevy2016} and observing their equilibrium behavior is challenging. But strongly-interacting Bose gases are also realized in optical lattices, where three-body losses are negligible, offering a clear path to study equilibrium many-body physics. Moreover the use of detection methods with single-atom resolution, as developed using metastable helium-4 ($^4$He$^*$) atoms \cite{schellekens2005, vassen2012, cayla2018}, is instrumental in reconstructing correlations to all orders \cite{dall2013, herce2023}. 

\begin{figure*}[ht!]
  \centering
  \includegraphics[width=\linewidth]{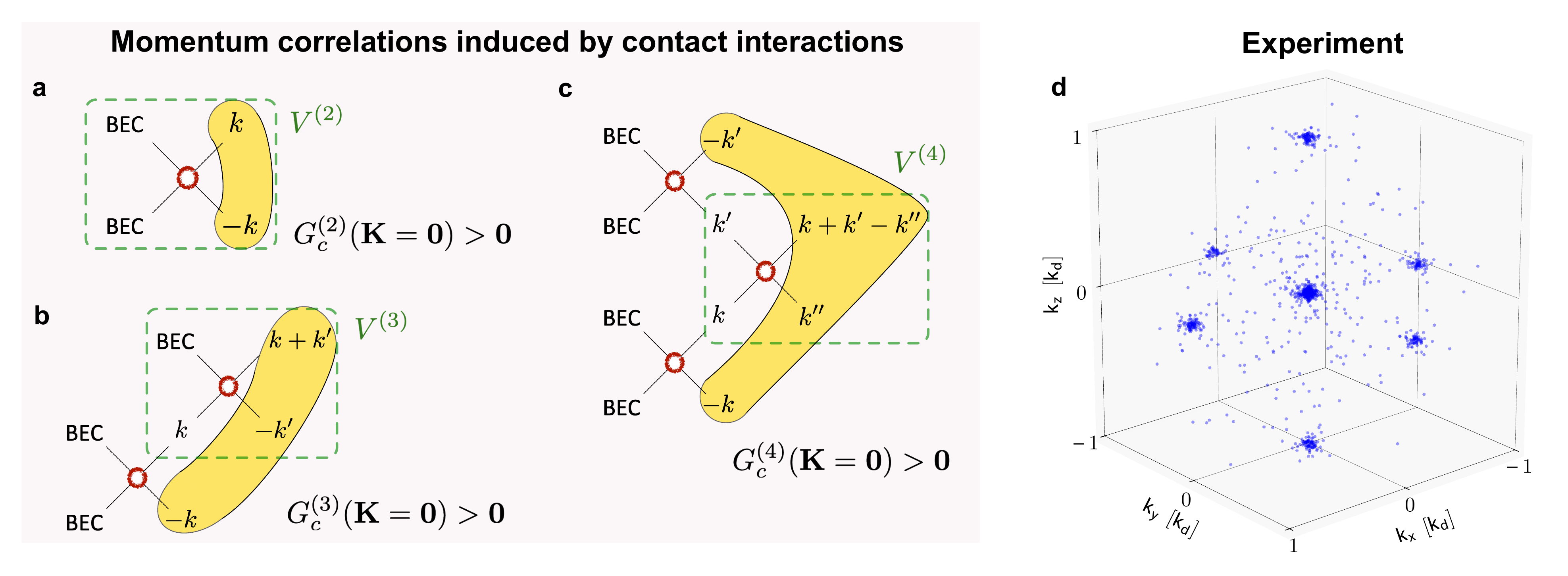}
  \caption{{\bf Two-body scattering processes from a Bose-Einstein condensate and momentum correlations}. 
  {\bf a} In Bogoliubov theory, the term $V^{(2)}$ in Eq.(\ref{Eq:Vbogo}) generates two-mode squeezed states at opposite momenta (yellow shaded area). {\bf b} Beyond the Bogoliubov approximation, the term $V^{(3)}$ in Eq.(\ref{Eq:Vtriplet}) involves the interaction of one atom in the BEC and one atom $\bm k \neq \bm0$. {\bf c} The beyond-Bogoliubov term $V^{(4)}$ in Eq.(\ref{Eq:Vquadrup}) describes the interaction of two atoms with $\bm k, \bm k' \neq \bm0$. At largest order, the terms $V^{(3)}$ and $V^{(4)}$ belong to series of scattering from the BEC, a framework from which triplets, quadruplets and $n$-uplets with ${\bm K}=\sum_{j=1}^n \bm k_{j}={\bm 0}$ emerge (yellow shaded areas). 
  {\bf d} Single-shot experimental distribution of individual atoms (blue dots) in momentum space, from which momentum correlations are computed.
  }
  \label{fig1}
\end{figure*}

As discussed above, the quantitative description of many-body correlations in strongly interacting Bose gases is generally a difficult problem. 
However, the mechanism by which interactions induce correlations in {\it momentum space} is rather intuitive. In a non-interacting Bose gas, assuming translational invariance, the populations $N_{\bm k}$ of each (quasi-)momentum mode ${\bm k}$ are independent of each other (up to the global constraint $\sum_{\bm k} N_{\bm k} = N$ with $N$ the total number of particles), and populations fluctuate incoherently (e.g. thermal fluctuations). When interactions are present, momentum modes are instead coupled and correlations between them emerge. The most elementary process of momentum-mode correlations comes from an elastic interaction between two particles, coupling two incoming modes $({\bm k}_1, {\bm k}_2)$ to two outgoing ones $({\bm k}_3, {\bm k}_4)$ which conserve the total momentum, ${\bm k}_1 +{\bm k}_2 = {\bm k}_3 + {\bm k}_4$. As we shall discuss below, this four-wave mixing process generates a hierarchy of subsets of correlated momentum modes.

To exploit momentum conservation in the four-wave mixing process, two-body contact interactions should be expressed in the momentum basis. The contact interaction term  $V_{\rm int}$ between bosons inside a volume $V$ reads
 $V_{\rm int}  = 
 \frac{U}{2 V} \sum_{{\bm k},{\bm k}',{\bm k}''} ( a^{\dagger}_{{\bm k}}a^{\dagger}_{{\bm k}'}a_{{\bm k}''}a_{{\bm k}+{\bm k}'-{\bm k}''} + {\rm h.c.} ) $
where $a_{\bm k}, a_{\bm k}^\dagger$ are destruction and creation operators for bosonic momentum mode $\bm k$, $U$ is the interaction strength, and ${\rm h.c.}$ indicates Hermitian conjugates. 
When interactions are weak and the temperature is close to zero, the state of the system is a nearly perfect condensate, namely $ N_{\bm k=0} = N_0 \approx N$. Then one can adopt the Bogoliubov replacement $a_0, a_0^\dagger \approx \sqrt{N_0}$, and rewrite the interactions as $V_{\rm int} \approx V^{(2)} + V^{(3)} + V^{(4)}$ \cite{FetterWalecka} where
\begin{eqnarray}
   V^{(2)}  &=  &\frac{U N_{0}^2}{2 V}  + \frac{U N_{0}}{2 V}  \sum_{{\bm k}\neq {\bm 0}}  \left ( a^{\dagger}_{{\bm k}}a^{\dagger}_{-{\bm k}} + {\rm h.c.}  \right ) \label{Eq:Vbogo} \\
   V^{(3)}  &=& \frac{U \sqrt{N_{0}}}{2 V} \sum_{{\bm k},{\bm k'}} \left ( \hat{a}^\dagger_{{\bm k + \bm k'}} \hat{a}^\dagger_{{- \bm k'}} \hat{a}_{{\bm k}} + {\rm h.c.} \right ) \label {Eq:Vtriplet} \\
   V^{(4)}  &=& \frac{U}{2 V} {\sum_{{\bm k},{\bm k}',{\bm k}''}} \left ( a^{\dagger}_{{\bm k}+{\bm k}'-{\bm k}''} a^{\dagger}_{{\bm k}''}a_{{\bm k}'}a_{{\bm k}} + {\rm h.c.} \right ) ~\label{Eq:Vquadrup}~.
\end{eqnarray}
Here the sums are restricted to momenta ${\bm k}, {\bm k'}, {\bm k''}$ such that only operators of modes with non-zero momentum are included. 
Bogoliubov theory, mentioned in the introduction, amounts to retaining only the quadratic term of Eq.~(\ref{Eq:Vbogo}) in the above expression, which corresponds to linearizing quantum fluctuations, and describing only pairwise correlations among opposite momenta. 

Moving beyond the weakly-interacting regime, quantum fluctuations become too large to be linearized, as the quantum depletion increases monotonically with the interaction strength.  
As a consequence the cubic and quartic terms, Eqs.~\eqref{Eq:Vtriplet} and \eqref{Eq:Vquadrup}, can no longer be neglected. Interestingly,  momentum correlations induced by these terms have a clear physical picture when considering successive scattering events, as illustrated in Fig.~\ref{fig1}b-c. The cubic term $V^{(3)}$ of Eq.~\eqref{Eq:Vtriplet} correlates triplets of modes whose momenta sum up to zero, $\bm K = \sum_{j=1}^3 \bm k_{j}= \bm 0$ while the quartic term $V^{(4)}$ of Eq.~\eqref{Eq:Vquadrup} correlates (among others) quartets of modes with $\bm K = \sum_{j=1}^4 \bm k_{j}=\bm 0$.

The emergence of correlated sets of $n>2$ modes, whose correlations are not reducible to pairwise ones, distinctly indicates a deviation from the Gaussian-state depiction. A quantitative understanding of this regime, and in particular of the structure of its correlations, remains challenging (see  e.g. Ref. \cite{carlen2021} for a recent theoretical effort in continuum space).  
To address this problem experimentally, we probe correlated sets of $n$ momentum modes (depicted in Fig.~\ref{fig1}a-c) via connected correlation functions $G_{\rm c}^{(n)}$ which are $n$-th order multivariate cumulants \cite{ursell1927}. Here, we probe connected correlations among $n$ momentum populations, $G_{\rm c}^{(n)}(N_{{\bm k}_1},N_{{\bm k}_2}, ..., N_{{\bm k}_n}) = \langle N_{{\bm k}_1} N_{{\bm k}_2} ... N_{{\bm k}_n} \rangle  - G_{\rm dis}^{(n)}(N_{{\bm k}_1},N_{{\bm k}_2}, ..., N_{{\bm k}_n})$ where the disconnected part $G_{\rm dis}^{(n)}$ includes the contributions of all lower-order correlations $n'<n$ to the total $n$-body correlations.  The $G_{\rm c}^{(n)}$ function isolates therefore genuine $n$-mode correlations, i.e. those not reducible to correlations of lower order. 
Connected correlations among zero-sum momentum modes, $G_{\rm c}^{(n)}(\bm K=0)$, represent the most relevant manifestation of quantum fluctuations induced by interactions, both at the level of Bogoliubov theory as well as beyond (Fig.~\ref{fig1}a-c). 
In contrast, incoherent (thermal) fluctuations are not subject to any momentum-sum constraint (as the momentum sum itself can fluctuate incoherently). Hence detecting correlations among zero-sum modes is a strong indication of their quantum origin, caused by processes of the kind depicted in Fig.~\ref{fig1}a-c. 
In the following, we shall focus our attention on the two-mode, zero-sum connected correlations $G_{\rm c}^{(2)}(\bm k, -\bm k)=\langle N_{\bm k} N_{-\bm k} \rangle  - \langle N_{\bm k} \rangle \langle N_{- \bm k} \rangle$. 
Bogoliubov theory predicts these correlations to take the form $G^{(2)}_{c}({\bm k},-{\bm k}) = | \langle a^{\dagger}_{\bm k} a^{\dagger}_{-\bm k}  \rangle |^2$ (Supplementary Information), namely as stemming from momentum pairing in the system. 

Our experiment begins with the production of a metastable helium-4 $\left( ^4 \textrm{He}^* \right)$ Bose-Einstein condensate (BEC) in a crossed optical dipole trap \cite{bouton2015}. The BEC is then adiabatically transferred to a 3D optical lattice \cite{carcy2021}, which we use to increase the interactions and the quantum depletion \cite{tenart2021}. In the lattice, the Bose gas is well described by the Bose-Hubbard Hamiltonian, $H = - J \sum_{\langle ij \rangle} (b_i^\dagger b_j + b_j^\dagger b_i) + V_{\rm int}$ with tunneling energy $J$ and on-site interaction energy $U$ -- here $\langle ij \rangle$ denotes a pair of nearest neighbors and $b_{j}^\dagger$ (resp. $b_{j}$) the creation (resp. destruction) operator on site $j$. The lattice wave vector is $k_d=2 \pi /d$ where the $d = 775$ nm is the lattice spacing. 
We access momentum correlations in a time-of-flight (TOF) experiment where the $^4$He$^*$ atoms fall onto a pair of Micro-Channel Plates (MCPs) located $43~$cm below the optical lattice. The combination of MCPs with delay lines \cite{nogrette2015} yields the 3D single-atom-resolved distribution of the gas in momentum space, as illustrated in Fig.~\ref{fig1}d. We use a total atom number $N= 5.0(7) \times 10^3$ to ensure that the lattice filling is unity in the center of the trap and that interactions are negligible during the TOF dynamics. Under these conditions, we are able to measure the in-trap momenta of individual atoms \cite{cayla2018,tenart2020}. 

\begin{figure*}[ht!]
  \centering
  \includegraphics[width=\linewidth]{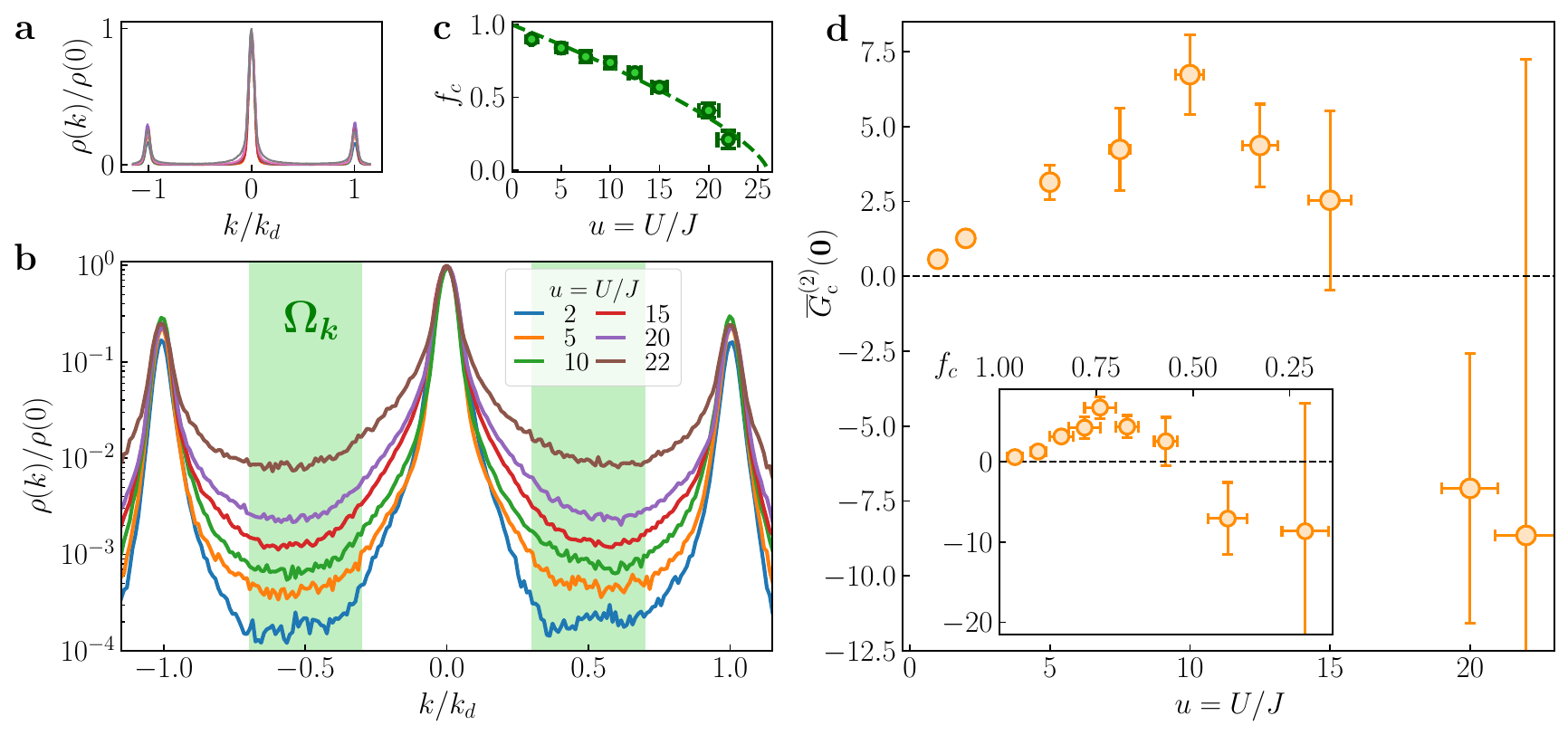}
  \caption{{\bf Momentum density profiles and two-body connected correlations at opposite momenta.} {\bf a} Normalised 1D cuts of the momentum density $\rho(k)/\rho(0)$ for varying ratios $u=U/J \in [2;22]$. {\bf b} Same as panel a in semi-log scale. The green shaded area represents the integration volume $\Omega_k$ of the depletion over which two-body correlations are computed. {\bf c} Condensate fraction $f_{c}$ as a function of $u=U/J$. The dashed line is a guide to the eye $\propto(1-u/u_{c})^{2 \beta}$ with $\beta=0.3485$ and $u_{c}=26$ \cite{herce2021}. Error bars reflect one standard deviation. {\bf d} Amplitude $\overline{G}_{\rm c}^{(2)}(\bm K = \bm 0)$ of the Bogoliubov momentum pairing at opposite momenta (${\bm K}={\bm k}_{1}+{\bm k}_{2}={\bm 0}$) measured in the volume $\Omega_{k}$ occupied by the depletion of the lattice BEC (orange circles). The measured values have been multiplied by $\eta^{-2}$ to account for the finite detection efficiency $\eta = 0.53(3)$ of our detector. The vertical error bars correspond to $68\%$-confidence intervals obtained from a bootstrap method. Inset: $\overline{G}_{\rm c}^{(2)}(\bm K = \bm 0)$ plotted as a function of the condensate fraction $f_{c}$.}
  \label{fig2}
\end{figure*}

We have recorded atom distributions at various interaction strengths $u=U/J$ in the range $u \in [1,22]$ ($\sim 2300$ distributions for each $u$),  far from the transition to the (finite-temperature) Mott insulator, occurring at the critical ratio $u_c = 26(1)$ in our experiment \cite{herce2021}. This is illustrated in Figure \ref{fig2}a, where normalised 1D cuts of the momentum density are shown to exhibit contrasted peaks associated to the presence of large condensate fractions. The semi-logarithmic scale in Figure \ref{fig2}b highlights the increase in the depletion density by more than one order of magnitude at the edge of the first Brillouin zone $k=0.5~k_{d}$. The corresponding condensate fraction $f_{c}$ is plotted as a function of interactions $u=U/J$ in Fig.~\ref{fig2}c. Importantly, we perform a thermometry \cite{carcy2021} and find that the change in the reduced temperature $T/J$ is compatible with an adiabatic variation with $U/J$ (Supplementary Information). The observed increase of the depletion in Fig.~\ref{fig2}a-c thus results predominantly from the increase of interactions (i.e. not an effect of heating). 
In the following we concentrate our attention on a region of momentum space, denoted by $\Omega_k$ (Fig.~\ref{fig2}b), which captures the quantum depletion of the condensate. In this work, we use for $\Omega_k$ a cubic corona comprised between two cubes centred on ${\bm k}={\bm 0}$ and of sides $2 \times 0.7~k_d$ and $2 \times 0.3~k_d$ respectively (Fig.~\ref{fig2}b and Methods). 

\begin{figure}
  \centering
   \includegraphics[width=\linewidth]{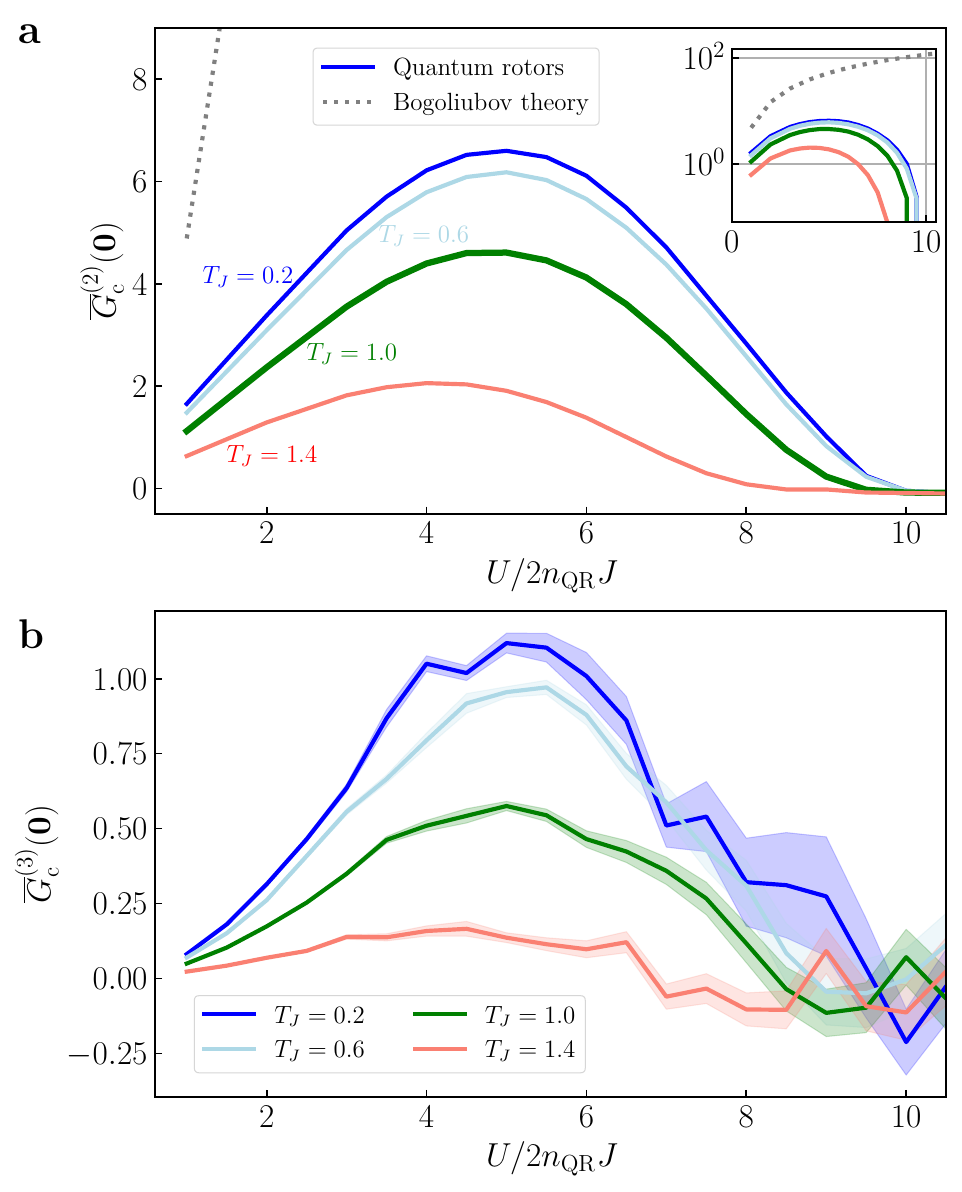}
  \caption{{\bf Numerical calculations of connected correlations $\overline{G}_{\rm c}^{(n)}( \bm 0)$ with the quantum rotor model.}
{\bf a} Two-mode connected correlations $\overline{G}_{\rm c}^{(2)}(\bm 0)$ obtained with the quantum rotor model (solid lines) at different temperatures, ranging from $T_{J}=k_{B} T/2n_{\rm QR}J=0.2$ (dark blue) to $T_{J}=1.4$ (dark red) in steps of 0.4. The system size is $10^3$. The dotted line is the result from Bogoliubov theory for the Bose-Hubbard model with $10^3$ sites and unity filling. Inset: log-plot of the data from the main panel. The amplitude of $\overline{G}_{\rm c}^{(2)}( \bm 0)$ is normalised to the measured atom number $\langle N_{\Omega_k} \rangle_{\rm exp}$  in the experiment (Methods and Supplementary Information).
{\bf b} Three-mode connected correlations $\overline{G}_{\rm c}^{(3)}( \bm 0)$ obtained with the quantum rotor model at the various temperatures. The amplitude of $\overline{G}_{\rm c}^{(3)}( \bm 0)$ is normalised to the measured atom number (Methods and Supplementary Information). The shaded area represents the uncertainty of one standard deviation.
}
  \label{fig3}
\end{figure}

From the experimental data, we extract the connected correlations at opposite momenta, $G_{\rm c}^{(2)}(\bm k,-\bm k)$, for momenta $\bm k$ belonging to $\Omega_k$.
To increase the signal-to-noise ratio, we sum $G_{\rm c}^{(2)}(\bm k,-\bm k)$ over $\Omega_k$, defining the integrated amplitude $\overline{G}_{\rm c}^{(2)}( \bm 0)= \sum_{\bm k \in \Omega_{k}} \ G_{\rm c}^{(2)}(\bm k,-\bm k)$.
In Figure \ref{fig2}d, we plot $\overline{G}_{\rm c}^{(2)}( \bm 0)$ as a function of the interaction strength $u$. 
We find that $\overline{G}_{\rm c}^{(2)}( \bm 0) \approx 0$ as $u \rightarrow 0$, confirming the absence of momentum pairing in non-interacting gases. Furthermore, $\overline{G}_{\rm c}^{(2)}( \bm 0)$ first rises as $u$ is increased from $u=0$. This observation is an excellent agreement with Bogoliubov theory, which predicts that the two-mode correlated quantum depletion always increases with $u$ (Fig.~\ref{fig3}a) \cite{toth2008}.
In contrast, our measurement shows that $\overline{G}_{\rm c}^{(2)}( \bm 0)$ decreases at larger values of $u$. The position of the maximum corresponds to a relatively large condensate fraction of $f_{c} \sim 0.75$ (inset of Fig.~\ref{fig2}d), far from the Mott transition. This strongly suggests that our observations are not specific to lattice bosons at integer filling, but are generic of strongly interacting Bose gases.
 
We stress that this non-monotonic variation of $\overline{G}_{\rm c}^{(2)}( \bm 0)$ results from an adiabatic change of the interaction strength \cite{carcy2021}, and is not the result of heating. 
It should also be contrasted with the variation of the bunching amplitude $g^{(2)}_{\rm HBT}(0) = \left ( \sum_{\bm k \in \Omega_k} \langle a^\dagger_{\bm k} a^\dagger_{\bm k} a_{\bm k} a_{\bm k} \rangle \right ) / \left ( \sum_{\bm k \in \Omega_k} \langle N_{\bm k} \rangle ^2 \right )$ set by the Bose statistics. Indeed, we find a constant and perfectly-contrasted value $g^{(2)}_{\rm HBT}(0) = 2.00(5)$ in all regimes of interactions (Supplementary Information). This result, expected at weak interactions, extends our previous observations \cite{cayla2020} to the strongly-interacting regime.
Finally, a further observation is the increase of the fluctuations of $\overline{G}_{\rm c}^{(2)}( \bm 0)$ with interactions shown by the error bars (see Fig.~\ref{fig2}d) obtained from a bootstrap analysis of the measured connected correlations.
This analysis suggests that the number of pairs at opposite momenta fluctuates more as interactions increase. An interesting question for a future work is to understand whether this behaviour stems from the increasing complexity of the equilibrium many-body state at strong interactions, made of correlated subsets with $n>2$ modes. 

To investigate the fate of two-mode connected correlations $\overline{G}_{\rm c}^{(2)}( \bm 0)$ beyond the weakly interacting regime theoretically, we make use of numerically exact quantum Monte Carlo (QMC) simulations on a 3D quantum rotor model (Supplementary Information), which captures the physics of the Bose-Hubbard model at large, integer filling $n_{\rm QR}$, and without confining potentials \cite{Wallinetal1994}. Calculations on a quantum-rotor model are not expected to be quantitatively accurate when compared with experiments conducted at nearly unit filling in a harmonic trap. 
Yet, unlike QMC simulations for the Bose-Hubbard model, they give access to arbitrary correlations in momentum space without the need for specialized update algorithms \cite{Roscilde2008, Fangetal2016}; and they do so across all regimes of temperatures and interactions up to the Mott transition (Supplementary Information).  
Fig.~\ref{fig3}a shows $\overline{G}_{\rm c}^{(2)}( \bm 0)$, integrated over quasi-momenta in a cubic corona $\Omega_{k}$ between two cubes of sides $2 \times 0.3 k_{d}$ and $2 \times 0.5 k_{d}$, for a system of $10^3$ cubic lattice sites, as a function of the interaction strength, and for different temperatures. The theoretical data have been normalised to the atom number $\langle N_{\Omega_k} \rangle$ detected experimentally at a given interaction strength $u$ (Methods). The temperature of the experiment was previously estimated to be around $k_B T/J \approx  2$ \cite{carcy2021}, corresponding to the data for the quantum rotor model at a temperature $T_{J}=k_B T/(2J n_{\rm QR}) = 1$.  

We observe that the quantum-rotor predictions capture the non-monotonic behavior of $\overline{G}_{\rm c}^{(2)}( \bm 0)$, as well as the right order of magnitude of the measured correlations. Our numerical results confirm that the decay of the connected two-body correlations $\overline{G}_{\rm c}^{(2)}( \bm 0)$ originate from quantum fluctuations beyond the linearized Bogoliubov regime. The maximum value of $\overline{G}_{\rm c}^{(2)}( \bm 0)$ can therefore be regarded as a microscopic signature of the Bose gas entering the strongly correlated regime, whose description requires accounting for correlations among $n>2$ momentum modes. 

To proceed with probing the hierarchy of momentum correlations illustrated in Fig.~\ref{fig1}, we studied the integrated  three-mode correlations $\overline{G}_{\rm c}^{(3)}( \bm 0) = \sum_{\bm k_1, \bm k_2} G^{(3)}_{\rm c}(\bm k_1, \bm k_2,  -\bm k_1 -\bm k_2)$ that reveal genuine three-mode correlations with zero momentum sum (Supplementary Information).
In the experiment, the measured values of $\overline{G}_{\rm c}^{(3)}(\bm 0)$ are consistent with zero for all the interaction strengths we probe. This result leaves the question open on the magnitude expected for three-mode correlations. We thus evaluate these correlations theoretically using the quantum-rotor model; and  in Fig.~\ref{fig3}b we plot $\overline{G}_{\rm c}^{(3)}( \bm 0)$, integrated over trios of modes in the cubic corona $\Omega_{k}$.
We observe that genuine three-mode correlations are positive, in agreement with the picture offered in Fig.~\ref{fig1}, and rise along with the two-mode correlations. This shows that the Bogoliubov picture of the Bose gas with intermediate interactions is in fact incomplete, and that significant non-Gaussian correlations are present in the system.
Nonetheless the magnitude of  $\overline{G}_{\rm c}^{(3)}( \bm 0)$ is significantly smaller than the one of $\overline{G}_{\rm c}^{(2)}( \bm 0)$, the temperature and interaction being the same. 
While two-mode correlations with total zero (quasi-)momentum uniquely identify the correlated pairs of modes, three-mode correlations with zero-sum (quasi-)momentum can couple every mode to $O(N)$ pairs of other modes, making correlations much weaker within each trio. Correlations are expected to be even more tenuous upon increasing the number $n$ of modes. Moreover, revealing triple coincidences of atoms in three modes is made difficult by the finite-detection efficiency $\eta$ of our detector. These aspects clearly make the observation of $n$-mode correlations with $n>2$ challenging. 
Finally, both forms of correlations (two-mode and three-mode) decrease upon moving to the strongly correlated regime; in the correlated superfluid regime they should scale linearly with the particle number, but they both become of $O(1)$ (i.e. not scaling with the system size) in the deep Mott limit, where they uniquely stem from particle-number conservation (Supplemental Material).

As we shall now explain, measuring $\overline{G}_{\rm c}^{(3)}(0)\neq0$ is, however, not the only means to demonstrate non-Gaussian statistics. 
In the spirit of Bogoliubov theory, we relax the assumption of particle-number conservation, but we consider that all the interaction processes described in Eqs.~(\ref{Eq:Vbogo})-(\ref{Eq:Vquadrup}) do not lead to the appearance of a finite value of $\langle a_{\bm k} \rangle$ in the depletion. This assumption is compatible with the perfectly-contrasted bunching, $g^{(2)}_{\rm HBT}(0) = 2.00(5)$, measured in the experiment as a finite value of $\langle a_{\bm k} \rangle$ would reduce the bunching amplitude. Hence, the connected two-mode correlations can then be expressed in terms of the 4-th-order cumulant $\langle a^{\dagger}_{\bm k}  a^{\dagger}_{-\bm k}  a_{\bm k}  a_{-\bm k}  \rangle_{c}$ (Supplemental Material),
\begin{align}
G^{(2)}_{c}({\bm k},-{\bm k})&= \nonumber \\  
| \langle a^{\dagger}_{\bm k}  &a^{\dagger}_{-\bm k}   \rangle |^2  + | \langle  a^{\dagger}_{\bm k}  a_{-\bm k}  \rangle |^2 
+  \langle a^{\dagger}_{\bm k}  a^{\dagger}_{-\bm k}  a_{\bm k}  a_{-\bm k}  \rangle_{c}. \label{Eq:ConnectCorr}
\end{align}
Equation~(\ref{Eq:ConnectCorr}) shows that negative values of $G^{(2)}_{c}({\bm k},-{\bm k})$ necessarily imply that $ \langle a^{\dagger}_{\bm k}  a^{\dagger}_{-\bm k}  a_{\bm k}  a_{-\bm k}  \rangle_{c} \neq 0$, which is a signature of non-Gaussian statistics.   
A similar approach to unveil a non-zero 4-th-order cumulant from the measurement of two-photon correlations has been used recently \cite{ferioli2024}.
A negative correlation $G_{\rm c}^{(2)}({\bm k},-{\bm k})<0$ amounts to  {\it anti-bunching}, namely the fact that the coincidence counts $\langle N_{\bm k} N_{-\bm k} \rangle$ is smaller than what expected in the absence of correlations, $\langle N_{\bm k} \rangle \langle N_{-\bm k}\rangle$.
Our numerical results on quantum rotors exhibit $G_{\rm c}^{(2)}({\bm k},-{\bm k})<0$ at large values $U/J$, when the homogeneous system enters the normal phase (Supplemental Material). This stems from the conservation of the total atom number, which indeed corresponds to non-Gaussian correlations as it involves all atoms.

In the experiment, we find a drastic change in the shape of $\overline{G}_{\rm c}^{(2)}(\delta {\bm k})$ at large interactions $u \geq 20$: there is hardly any correlation peak at $\delta {\bm k}={\bm 0}$ but a constant (negative) offset independent of $\delta {\bm k}$ (Supplemental Material). This anti-bunching thus differs from that of two fermions which takes place at short separations \cite{Rom2006}: the absence of a characteristic two-body length scale suggests that the observed $\overline{G}_{\rm c}^{(2)}({\bm 0})<0$ is set by correlations between more than two atoms in momentum space.
As discussed above, $\overline{G}_{\rm c}^{(2)}( \bm 0)<0$  implies $\sum_{{\bm k}} \langle a^{\dagger}_{\bm k}  a^{\dagger}_{-\bm k}  a_{\bm k}  a_{-\bm k}  \rangle_{c} \neq 0$, which is an unambiguous signature of non-Gaussian statistics. Importantly, this signature  does not rely on any theoretical modeling. As such, it represents a striking evidence of the strongly correlated (non-Gaussian) nature of Bose gases at large interactions. Determining whether the observed $\overline{G}_{\rm c}^{(2)}({\bm 0})<0$ in the harmonically-trapped gas shares the same physical origin as that found in numerical studies of homogeneous systems is a complex question that requires further investigation.

In conclusion, by making use of state-of-the-art preparation and detection techniques for ultracold gases, and of advanced numerical methods, we provide a physical picture of correlations in an interacting Bose gas beyond the Bogoliubov regime. Two-mode correlations, at the heart of the Bogoliubov weakly interacting regime, are suppressed when entering the strongly interacting regime, in favor of $n$-mode correlations with $n>2$, and non-Gaussian correlations appear. These results suggest  that, upon increasing the interactions,  entanglement in momentum space (that reflects itself in the correlations we probed in this work) becomes increasingly complex in its multipartite nature; as well as also increasingly elusive, since each mode becomes weakly entangled with a myriad (${\cal O}(N^{n-2})$) of $(n-1)$-plets  of other modes to form $n$-mode entangled clusters. Finding a proper theoretical characterization, as well as experimental signatures, of this complex entanglement structure represents a fascinating open question  for future studies.  

\section*{Acknowledgements}

We thank E. Gradova for her contribution to the bootstrap analysis, T. Chalopin for a careful reading of the manuscript, D. Boiron and the members of the Quantum Gas group at Institut d'Optique for insightful discussions. We acknowledge financial support from the R\'egion Ile-de-France in the framework of the DIM SIRTEQ, the ``Fondation d'entreprise iXcore pour la Recherche", the French National Research Agency (Grant number ANR-17-CE30-0020-01) and France 2030 programs of the French National Research Agency (Grant numbers ANR-22-PETQ-0004 and ANR-11-IDEX-0003). All numerical simulations were made on the PSMN cluster at the ENS of Lyon.

\section*{Methods}
{\it Measurement of connected correlations at opposite momenta.} The two-body correlations shown in the main text are calculated in the depletion of the condensates, considering atoms belonging to the volume $\Omega_{k}$ defined as the volume between two cubes of sides $2 \times 0.3~k_d$ and $2 \times 0.7~k_d$. Since $\Omega_k$ corresponds to $1.8 \leq |k| \xi \leq 3.0$ at $U/J=2$ and $1.2 \leq |k| \xi \leq 2.0$ at $U/J=22$, with $\xi$ the healing length, it excludes by design the phononic part $\left( |k| \xi \ll 1 \right)$ of the thermal depletion. Two-body correlations at opposite momenta $G_c^{(2)}$ are computed from {\it (i)} calculating the histogram of momentum sums ${\bm k}_{1}+{\bm k}_{2}$ for all pairs of atoms detected in $\Omega_{k}$ in a run of the experiment \cite{tenart2021} (from which we subtract the product of average occupations $\langle N(\bm k_{1}) \rangle \langle N(\bm k_{2}) \rangle$), and {\it (ii)} integrating this histogram over a cubic volume centered on ${\bm k}_{1}+{\bm k}_{2}={\bm 0}$ and of size $0.18~k_d$ set by the width of the two-body correlation function. 
\\
 
{\it Comparison of numerical results from quantum rotor model with experiments.} Connected correlations $\overline{G}_{\rm c}^{(2)}(0)$ and $\overline{G}_{\rm c}^{(3)}(0)$ are computed numerically with the quantum rotor model over a cubic corona $\Omega_{k}$ between two cubes of sides $2 \times 0.3 ~k_{d}$ and $2 \times 0.5~k_{d}$. These calculations are performed with a large, integer filling $n_{\rm QR}$ of the 3D lattice (Supplemental material), which leaves us with an adjustable parameter to compare with experiments (performed in the Bose-Hubbard regime with one atom per site at the trap center). The numerically-computed population $\langle N_{\Omega_{k}}^{th} \rangle$ of the modes in momentum space changes with the filling $n_{\rm QR}$. To plot the numerical results of $\overline{G}_{\rm c}^{(2)}(0)$ ({\it resp.} $\overline{G}_{\rm c}^{(3)}(0)$) in the main text, we have rescaled their amplitude by $(\langle N_{\Omega_{k}}\rangle / \langle N_{\Omega_{k}}^{th} \rangle )^2$ ({\it resp.} $(\langle N_{\Omega_{k}}\rangle / \langle N_{\Omega_{k}}^{th} \rangle )^3$), where $\langle N_{\Omega_{k}}\rangle$ is the measured momentum population observed in the experiment.
 
%%%%%%%%%%%%%%%%%%%%%%%%%%%%%%%%%%%%%%%
\bibliography{Bureik2023-biblio.bib}

\begin{thebibliography}{62}%
\makeatletter
\providecommand \@ifxundefined [1]{%
 \@ifx{#1\undefined}
}%
\providecommand \@ifnum [1]{%
 \ifnum #1\expandafter \@firstoftwo
 \else \expandafter \@secondoftwo
 \fi
}%
\providecommand \@ifx [1]{%
 \ifx #1\expandafter \@firstoftwo
 \else \expandafter \@secondoftwo
 \fi
}%
\providecommand \natexlab [1]{#1}%
\providecommand \enquote  [1]{``#1''}%
\providecommand \bibnamefont  [1]{#1}%
\providecommand \bibfnamefont [1]{#1}%
\providecommand \citenamefont [1]{#1}%
\providecommand \href@noop [0]{\@secondoftwo}%
\providecommand \href [0]{\begingroup \@sanitize@url \@href}%
\providecommand \@href[1]{\@@startlink{#1}\@@href}%
\providecommand \@@href[1]{\endgroup#1\@@endlink}%
\providecommand \@sanitize@url [0]{\catcode `\\12\catcode `\$12\catcode
  `\&12\catcode `\#12\catcode `\^12\catcode `\_12\catcode `\%12\relax}%
\providecommand \@@startlink[1]{}%
\providecommand \@@endlink[0]{}%
\providecommand \url  [0]{\begingroup\@sanitize@url \@url }%
\providecommand \@url [1]{\endgroup\@href {#1}{\urlprefix }}%
\providecommand \urlprefix  [0]{URL }%
\providecommand \Eprint [0]{\href }%
\providecommand \doibase [0]{http://dx.doi.org/}%
\providecommand \selectlanguage [0]{\@gobble}%
\providecommand \bibinfo  [0]{\@secondoftwo}%
\providecommand \bibfield  [0]{\@secondoftwo}%
\providecommand \translation [1]{[#1]}%
\providecommand \BibitemOpen [0]{}%
\providecommand \bibitemStop [0]{}%
\providecommand \bibitemNoStop [0]{.\EOS\space}%
\providecommand \EOS [0]{\spacefactor3000\relax}%
\providecommand \BibitemShut  [1]{\csname bibitem#1\endcsname}%
\let\auto@bib@innerbib\@empty
%</preamble>
\bibitem [{\citenamefont {Sachdev}(2023)}]{Sachdevbook2023}%
  \BibitemOpen
  \bibfield  {author} {\bibinfo {author} {Sachdev, S.},\ }\href@noop {} {\emph
  {\bibinfo {title} {Quantum Phases of Matter}}},\ \bibinfo {edition} {1st}\
  ed.\ (\bibinfo  {publisher} {Cambridge University Press},\ \bibinfo {year}
  {2023})\BibitemShut {NoStop}%
\bibitem [{\citenamefont {Morosan}\ \emph {et~al.}(2012)\citenamefont
  {Morosan}, \citenamefont {Natelson}, \citenamefont {Nevidomskyy},\ and\
  \citenamefont {Si}}]{Morosan2012}%
  \BibitemOpen
  \bibfield  {author} {\bibinfo {author} {Morosan, E.}, \bibinfo {author}
  {Natelson, D.}, \bibinfo {author} {Nevidomskyy, A.~H.}\ and\ \bibinfo
  {author} {Si, Q.},\ }\bibfield  {title} {\bibinfo {title} {\emph {Strongly
  Correlated Materials}},\ }\href {\doibase
  https://doi.org/10.1002/adma.201202018} {\bibfield  {journal} {\bibinfo
  {journal} {Advanced Materials}\ }\textbf {\bibinfo {volume} {24}},\ \bibinfo
  {pages} {4896} (\bibinfo {year} {2012})}\BibitemShut {NoStop}%
\bibitem [{\citenamefont {Adams}\ \emph {et~al.}(2012)\citenamefont {Adams},
  \citenamefont {Carr}, \citenamefont {Schäfer}, \citenamefont {Steinberg},\
  and\ \citenamefont {Thomas}}]{Adams2012}%
  \BibitemOpen
  \bibfield  {author} {\bibinfo {author} {Adams, A.}, \bibinfo {author} {Carr,
  L.~D.}, \bibinfo {author} {Schäfer, T.}, \bibinfo {author} {Steinberg, P.}\
  and\ \bibinfo {author} {Thomas, J.~E.},\ }\bibfield  {title} {\bibinfo
  {title} {\emph {Strongly correlated quantum fluids: ultracold quantum gases,
  quantum chromodynamic plasmas and holographic duality}},\ }\href {\doibase
  10.1088/1367-2630/14/11/115009} {\bibfield  {journal} {\bibinfo  {journal}
  {New Journal of Physics}\ }\textbf {\bibinfo {volume} {14}},\ \bibinfo
  {pages} {115009} (\bibinfo {year} {2012})}\BibitemShut {NoStop}%
\bibitem [{\citenamefont {Bloch}\ \emph {et~al.}(2022)\citenamefont {Bloch},
  \citenamefont {Cavalleri}, \citenamefont {Galitski}, \citenamefont {Hafezi},\
  and\ \citenamefont {Rubio}}]{Bloch2022}%
  \BibitemOpen
  \bibfield  {author} {\bibinfo {author} {Bloch, J.}, \bibinfo {author}
  {Cavalleri, A.}, \bibinfo {author} {Galitski, V.}, \bibinfo {author} {Hafezi,
  M.}\ and\ \bibinfo {author} {Rubio, A.},\ }\bibfield  {title} {\bibinfo
  {title} {\emph {Strongly correlated electron--photon systems}},\ }\href
  {\doibase 10.1038/s41586-022-04726-w} {\bibfield  {journal} {\bibinfo
  {journal} {Nature}\ }\textbf {\bibinfo {volume} {606}},\ \bibinfo {pages}
  {41} (\bibinfo {year} {2022})}\BibitemShut {NoStop}%
\bibitem [{\citenamefont {Chevy}\ and\ \citenamefont
  {Salomon}(2016)}]{Chevy2016}%
  \BibitemOpen
  \bibfield  {author} {\bibinfo {author} {Chevy, F.}\ and\ \bibinfo {author}
  {Salomon, C.},\ }\bibfield  {title} {\bibinfo {title} {\emph {Strongly
  correlated Bose gases}},\ }\href {\doibase 10.1088/0953-4075/49/19/192001}
  {\bibfield  {journal} {\bibinfo  {journal} {Journal of Physics B: Atomic,
  Molecular and Optical Physics}\ }\textbf {\bibinfo {volume} {49}},\ \bibinfo
  {pages} {192001} (\bibinfo {year} {2016})}\BibitemShut {NoStop}%
\bibitem [{\citenamefont {Georges}\ \emph {et~al.}(1996)\citenamefont
  {Georges}, \citenamefont {Kotliar}, \citenamefont {Krauth},\ and\
  \citenamefont {Rozenberg}}]{georges1996}%
  \BibitemOpen
  \bibfield  {author} {\bibinfo {author} {Georges, A.}, \bibinfo {author}
  {Kotliar, G.}, \bibinfo {author} {Krauth, W.}\ and\ \bibinfo {author}
  {Rozenberg, M.~J.},\ }\bibfield  {title} {\bibinfo {title} {\emph {Dynamical
  mean-field theory of strongly correlated fermion systems and the limit of
  infinite dimensions}},\ }\href {\doibase 10.1103/RevModPhys.68.13} {\bibfield
   {journal} {\bibinfo  {journal} {Rev. Mod. Phys.}\ }\textbf {\bibinfo
  {volume} {68}},\ \bibinfo {pages} {13} (\bibinfo {year} {1996})}\BibitemShut
  {NoStop}%
\bibitem [{\citenamefont {Metzner}\ \emph {et~al.}(2012)\citenamefont
  {Metzner}, \citenamefont {Salmhofer}, \citenamefont {Honerkamp},
  \citenamefont {Meden},\ and\ \citenamefont {Sch\"onhammer}}]{metzner2012}%
  \BibitemOpen
  \bibfield  {author} {\bibinfo {author} {Metzner, W.}, \bibinfo {author}
  {Salmhofer, M.}, \bibinfo {author} {Honerkamp, C.}, \bibinfo {author} {Meden,
  V.}\ and\ \bibinfo {author} {Sch\"onhammer, K.},\ }\bibfield  {title}
  {\bibinfo {title} {\emph {Functional renormalization group approach to
  correlated fermion systems}},\ }\href {\doibase 10.1103/RevModPhys.84.299}
  {\bibfield  {journal} {\bibinfo  {journal} {Rev. Mod. Phys.}\ }\textbf
  {\bibinfo {volume} {84}},\ \bibinfo {pages} {299} (\bibinfo {year}
  {2012})}\BibitemShut {NoStop}%
\bibitem [{\citenamefont {Georges}\ \emph {et~al.}(2013)\citenamefont
  {Georges}, \citenamefont {Medici},\ and\ \citenamefont
  {Mravlje}}]{Georges2013}%
  \BibitemOpen
  \bibfield  {author} {\bibinfo {author} {Georges, A.}, \bibinfo {author}
  {Medici, L.~d.}\ and\ \bibinfo {author} {Mravlje, J.},\ }\bibfield  {title}
  {\bibinfo {title} {\emph {Strong Correlations from Hund’s Coupling}},\
  }\href {\doibase 10.1146/annurev-conmatphys-020911-125045} {\bibfield
  {journal} {\bibinfo  {journal} {Annual Review of Condensed Matter Physics}\
  }\textbf {\bibinfo {volume} {4}},\ \bibinfo {pages} {137} (\bibinfo {year}
  {2013})}\BibitemShut {NoStop}%
\bibitem [{\citenamefont {Chang}\ \emph {et~al.}(2014)\citenamefont {Chang},
  \citenamefont {Vuleti{\'{c}}},\ and\ \citenamefont {Lukin}}]{Chang2014}%
  \BibitemOpen
  \bibfield  {author} {\bibinfo {author} {Chang, D.~E.}, \bibinfo {author}
  {Vuleti{\'{c}}, V.}\ and\ \bibinfo {author} {Lukin, M.~D.},\ }\bibfield
  {title} {\bibinfo {title} {\emph {Quantum nonlinear optics --- photon by
  photon}},\ }\href {\doibase 10.1038/nphoton.2014.192} {\bibfield  {journal}
  {\bibinfo  {journal} {Nature Photonics}\ }\textbf {\bibinfo {volume} {8}},\
  \bibinfo {pages} {685} (\bibinfo {year} {2014})}\BibitemShut {NoStop}%
\bibitem [{\citenamefont {Pitaevskii}\ and\ \citenamefont
  {Stringari}(2016)}]{Stringaribook}%
  \BibitemOpen
  \bibfield  {author} {\bibinfo {author} {Pitaevskii, L.}\ and\ \bibinfo
  {author} {Stringari, S.},\ }\href {\doibase
  10.1093/acprof:oso/9780198758884.001.0001} {\emph {\bibinfo {title}
  {{Bose-Einstein Condensation and Superfluidity}}}}\ (\bibinfo  {publisher}
  {Oxford University Press},\ \bibinfo {year} {2016})\BibitemShut {NoStop}%
\bibitem [{\citenamefont {Leggett}(2006)}]{leggettbook}%
  \BibitemOpen
  \bibfield  {author} {\bibinfo {author} {Leggett, A.},\ }\href
  {https://books.google.fr/books?id=kywSDAAAQBAJ} {\emph {\bibinfo {title}
  {Quantum Liquids: Bose Condensation and Cooper Pairing in Condensed-matter
  Systems}}},\ Oxford Graduate Texts\ (\bibinfo  {publisher} {OUP Oxford},\
  \bibinfo {year} {2006})\BibitemShut {NoStop}%
\bibitem [{\citenamefont {Makotyn}\ \emph {et~al.}(2014)\citenamefont
  {Makotyn}, \citenamefont {Klauss}, \citenamefont {Goldberger}, \citenamefont
  {Cornell},\ and\ \citenamefont {Jin}}]{Makotyn2014}%
  \BibitemOpen
  \bibfield  {author} {\bibinfo {author} {Makotyn, P.}, \bibinfo {author}
  {Klauss, C.~E.}, \bibinfo {author} {Goldberger, D.~L.}, \bibinfo {author}
  {Cornell, E.~A.}\ and\ \bibinfo {author} {Jin, D.~S.},\ }\bibfield  {title}
  {\bibinfo {title} {\emph {Universal dynamics of a degenerate unitary Bose
  gas}},\ }\href {\doibase 10.1038/nphys2850} {\bibfield  {journal} {\bibinfo
  {journal} {Nature Physics}\ }\textbf {\bibinfo {volume} {10}},\ \bibinfo
  {pages} {116} (\bibinfo {year} {2014})}\BibitemShut {NoStop}%
\bibitem [{\citenamefont {Fletcher}\ \emph {et~al.}(2017)\citenamefont
  {Fletcher}, \citenamefont {Lopes}, \citenamefont {Man}, \citenamefont
  {Navon}, \citenamefont {Smith}, \citenamefont {Zwierlein},\ and\
  \citenamefont {Hadzibabic}}]{Fletcher2016}%
  \BibitemOpen
  \bibfield  {author} {\bibinfo {author} {Fletcher, R.~J.}, \bibinfo {author}
  {Lopes, R.}, \bibinfo {author} {Man, J.}, \bibinfo {author} {Navon, N.},
  \bibinfo {author} {Smith, R.~P.}, \bibinfo {author} {Zwierlein, M.~W.}\ and\
  \bibinfo {author} {Hadzibabic, Z.},\ }\bibfield  {title} {\bibinfo {title}
  {\emph {Two- and three-body contacts in the unitary Bose gas}},\ }\href
  {\doibase 10.1126/science.aai8195} {\bibfield  {journal} {\bibinfo  {journal}
  {Science}\ }\textbf {\bibinfo {volume} {355}},\ \bibinfo {pages} {377}
  (\bibinfo {year} {2017})}\BibitemShut {NoStop}%
\bibitem [{\citenamefont {Eismann}\ \emph {et~al.}(2016)\citenamefont
  {Eismann}, \citenamefont {Khaykovich}, \citenamefont {Laurent}, \citenamefont
  {Ferrier-Barbut}, \citenamefont {Rem}, \citenamefont {Grier}, \citenamefont
  {Delehaye}, \citenamefont {Chevy}, \citenamefont {Salomon}, \citenamefont
  {Ha},\ and\ \citenamefont {Chin}}]{eismann2016}%
  \BibitemOpen
  \bibfield  {author} {\bibinfo {author} {Eismann, U.} \emph {et~al.},\
  }\bibfield  {title} {\bibinfo {title} {\emph {Universal Loss Dynamics in a
  Unitary Bose Gas}},\ }\href {\doibase 10.1103/PhysRevX.6.021025} {\bibfield
  {journal} {\bibinfo  {journal} {Phys. Rev. X}\ }\textbf {\bibinfo {volume}
  {6}},\ \bibinfo {pages} {021025} (\bibinfo {year} {2016})}\BibitemShut
  {NoStop}%
\bibitem [{\citenamefont {Klauss}\ \emph {et~al.}(2017)\citenamefont {Klauss},
  \citenamefont {Xie}, \citenamefont {Lopez-Abadia}, \citenamefont {D'Incao},
  \citenamefont {Hadzibabic}, \citenamefont {Jin},\ and\ \citenamefont
  {Cornell}}]{Klauss2017}%
  \BibitemOpen
  \bibfield  {author} {\bibinfo {author} {Klauss, C.~E.}, \bibinfo {author}
  {Xie, X.}, \bibinfo {author} {Lopez-Abadia, C.}, \bibinfo {author} {D'Incao,
  J.~P.}, \bibinfo {author} {Hadzibabic, Z.}, \bibinfo {author} {Jin, D.~S.}\
  and\ \bibinfo {author} {Cornell, E.~A.},\ }\bibfield  {title} {\bibinfo
  {title} {\emph {Observation of Efimov Molecules Created from a Resonantly
  Interacting Bose Gas}},\ }\href {\doibase 10.1103/PhysRevLett.119.143401}
  {\bibfield  {journal} {\bibinfo  {journal} {Phys. Rev. Lett.}\ }\textbf
  {\bibinfo {volume} {119}},\ \bibinfo {pages} {143401} (\bibinfo {year}
  {2017})}\BibitemShut {NoStop}%
\bibitem [{\citenamefont {Eigen}\ \emph {et~al.}(2018)\citenamefont {Eigen},
  \citenamefont {Glidden}, \citenamefont {Lopes}, \citenamefont {Cornell},
  \citenamefont {Smith},\ and\ \citenamefont {Hadzibabic}}]{Eigen2018}%
  \BibitemOpen
  \bibfield  {author} {\bibinfo {author} {Eigen, C.}, \bibinfo {author}
  {Glidden, J. A.~P.}, \bibinfo {author} {Lopes, R.}, \bibinfo {author}
  {Cornell, E.~A.}, \bibinfo {author} {Smith, R.~P.}\ and\ \bibinfo {author}
  {Hadzibabic, Z.},\ }\bibfield  {title} {\bibinfo {title} {\emph {Universal
  prethermal dynamics of Bose gases quenched to unitarity}},\ }\href {\doibase
  10.1038/s41586-018-0674-1} {\bibfield  {journal} {\bibinfo  {journal}
  {Nature}\ }\textbf {\bibinfo {volume} {563}},\ \bibinfo {pages} {221}
  (\bibinfo {year} {2018})}\BibitemShut {NoStop}%
\bibitem [{\citenamefont {Yan}\ \emph {et~al.}(2020)\citenamefont {Yan},
  \citenamefont {Ni}, \citenamefont {Robens},\ and\ \citenamefont
  {Zwierlein}}]{Yan2020}%
  \BibitemOpen
  \bibfield  {author} {\bibinfo {author} {Yan, Z.~Z.}, \bibinfo {author} {Ni,
  Y.}, \bibinfo {author} {Robens, C.}\ and\ \bibinfo {author} {Zwierlein,
  M.~W.},\ }\bibfield  {title} {\bibinfo {title} {\emph {Bose polarons near
  quantum criticality}},\ }\href {\doibase 10.1126/science.aax5850} {\bibfield
  {journal} {\bibinfo  {journal} {Science}\ }\textbf {\bibinfo {volume}
  {368}},\ \bibinfo {pages} {190} (\bibinfo {year} {2020})}\BibitemShut
  {NoStop}%
\bibitem [{\citenamefont {Bogoliubov}(1947)}]{Bogoliubov1947}%
  \BibitemOpen
  \bibfield  {author} {\bibinfo {author} {Bogoliubov, N.},\ }\bibfield  {title}
  {\bibinfo {title} {\emph {On the theory of superfluidity}},\ }\href@noop {}
  {\bibfield  {journal} {\bibinfo  {journal} {Journal of Physics (USSR)}\
  }\textbf {\bibinfo {volume} {11}},\ \bibinfo {pages} {23} (\bibinfo {year}
  {1947})}\BibitemShut {NoStop}%
\bibitem [{\citenamefont {Tenart}\ \emph {et~al.}(2021)\citenamefont {Tenart},
  \citenamefont {Herc{\'e}}, \citenamefont {Bureik}, \citenamefont {Dareau},\
  and\ \citenamefont {Cl{\'e}ment}}]{tenart2021}%
  \BibitemOpen
  \bibfield  {author} {\bibinfo {author} {Tenart, A.}, \bibinfo {author}
  {Herc{\'e}, G.}, \bibinfo {author} {Bureik, J.-P.}, \bibinfo {author}
  {Dareau, A.}\ and\ \bibinfo {author} {Cl{\'e}ment, D.},\ }\bibfield  {title}
  {\bibinfo {title} {\emph {Observation of pairs of atoms at opposite momenta
  in an equilibrium interacting Bose gas}},\ }\href {\doibase
  10.1038/s41567-021-01381-2} {\bibfield  {journal} {\bibinfo  {journal}
  {Nature Physics}\ }\textbf {\bibinfo {volume} {17}},\ \bibinfo {pages} {1364}
  (\bibinfo {year} {2021})}\BibitemShut {NoStop}%
\bibitem [{\citenamefont {Paschen}\ and\ \citenamefont
  {Si}(2021)}]{Paschen2021}%
  \BibitemOpen
  \bibfield  {author} {\bibinfo {author} {Paschen, S.}\ and\ \bibinfo {author}
  {Si, Q.},\ }\bibfield  {title} {\bibinfo {title} {\emph {Quantum phases
  driven by strong correlations}},\ }\href {\doibase
  10.1038/s42254-020-00262-6} {\bibfield  {journal} {\bibinfo  {journal}
  {Nature Reviews Physics}\ }\textbf {\bibinfo {volume} {3}},\ \bibinfo {pages}
  {9} (\bibinfo {year} {2021})}\BibitemShut {NoStop}%
\bibitem [{\citenamefont {Polkovnikov}\ \emph {et~al.}(2011)\citenamefont
  {Polkovnikov}, \citenamefont {Sengupta}, \citenamefont {Silva},\ and\
  \citenamefont {Vengalattore}}]{Polkolnikov2011}%
  \BibitemOpen
  \bibfield  {author} {\bibinfo {author} {Polkovnikov, A.}, \bibinfo {author}
  {Sengupta, K.}, \bibinfo {author} {Silva, A.}\ and\ \bibinfo {author}
  {Vengalattore, M.},\ }\bibfield  {title} {\bibinfo {title} {\emph
  {Colloquium: Nonequilibrium dynamics of closed interacting quantum
  systems}},\ }\href {\doibase 10.1103/RevModPhys.83.863} {\bibfield  {journal}
  {\bibinfo  {journal} {Rev. Mod. Phys.}\ }\textbf {\bibinfo {volume} {83}},\
  \bibinfo {pages} {863} (\bibinfo {year} {2011})}\BibitemShut {NoStop}%
\bibitem [{\citenamefont {Wang}\ \emph {et~al.}(2007)\citenamefont {Wang},
  \citenamefont {Hiroshima}, \citenamefont {Tomita},\ and\ \citenamefont
  {Hayashi}}]{Gaussian_states}%
  \BibitemOpen
  \bibfield  {author} {\bibinfo {author} {Wang, X.-B.}, \bibinfo {author}
  {Hiroshima, T.}, \bibinfo {author} {Tomita, A.}\ and\ \bibinfo {author}
  {Hayashi, M.},\ }\bibfield  {title} {\bibinfo {title} {\emph {Quantum
  information with Gaussian states}},\ }\href {\doibase
  https://doi.org/10.1016/j.physrep.2007.04.005} {\bibfield  {journal}
  {\bibinfo  {journal} {Physics Reports}\ }\textbf {\bibinfo {volume} {448}},\
  \bibinfo {pages} {1} (\bibinfo {year} {2007})}\BibitemShut {NoStop}%
\bibitem [{\citenamefont {Armijo}\ \emph {et~al.}(2010)\citenamefont {Armijo},
  \citenamefont {Jacqmin}, \citenamefont {Kheruntsyan},\ and\ \citenamefont
  {Bouchoule}}]{Armijo2010}%
  \BibitemOpen
  \bibfield  {author} {\bibinfo {author} {Armijo, J.}, \bibinfo {author}
  {Jacqmin, T.}, \bibinfo {author} {Kheruntsyan, K.~V.}\ and\ \bibinfo {author}
  {Bouchoule, I.},\ }\bibfield  {title} {\bibinfo {title} {\emph {Probing
  Three-Body Correlations in a Quantum Gas Using the Measurement of the Third
  Moment of Density Fluctuations}},\ }\href {\doibase
  10.1103/PhysRevLett.105.230402} {\bibfield  {journal} {\bibinfo  {journal}
  {Phys. Rev. Lett.}\ }\textbf {\bibinfo {volume} {105}},\ \bibinfo {pages}
  {230402} (\bibinfo {year} {2010})}\BibitemShut {NoStop}%
\bibitem [{\citenamefont {Schweigler}\ \emph {et~al.}(2017)\citenamefont
  {Schweigler}, \citenamefont {Kasper}, \citenamefont {Erne}, \citenamefont
  {Mazets}, \citenamefont {Rauer}, \citenamefont {Cataldini}, \citenamefont
  {Langen}, \citenamefont {Gasenzer}, \citenamefont {Berges},\ and\
  \citenamefont {Schmiedmayer}}]{Schweigler2017}%
  \BibitemOpen
  \bibfield  {author} {\bibinfo {author} {Schweigler, T.}, \bibinfo {author}
  {Kasper, V.}, \bibinfo {author} {Erne, S.}, \bibinfo {author} {Mazets, I.},
  \bibinfo {author} {Rauer, B.}, \bibinfo {author} {Cataldini, F.}, \bibinfo
  {author} {Langen, T.}, \bibinfo {author} {Gasenzer, T.}, \bibinfo {author}
  {Berges, J.}\ and\ \bibinfo {author} {Schmiedmayer, J.},\ }\bibfield  {title}
  {\bibinfo {title} {\emph {Experimental characterization of a quantum
  many-body system via higher-order correlations}},\ }\href@noop {} {\bibfield
  {journal} {\bibinfo  {journal} {Nature}\ }\textbf {\bibinfo {volume} {545}},\
  \bibinfo {pages} {323} (\bibinfo {year} {2017})}\BibitemShut {NoStop}%
\bibitem [{\citenamefont {Rispoli}\ \emph {et~al.}(2019)\citenamefont
  {Rispoli}, \citenamefont {Lukin}, \citenamefont {Schittko}, \citenamefont
  {Kim}, \citenamefont {Tai}, \citenamefont {L{\'e}onard},\ and\ \citenamefont
  {Greiner}}]{Rispoli2019}%
  \BibitemOpen
  \bibfield  {author} {\bibinfo {author} {Rispoli, M.}, \bibinfo {author}
  {Lukin, A.}, \bibinfo {author} {Schittko, R.}, \bibinfo {author} {Kim, S.},
  \bibinfo {author} {Tai, M.~E.}, \bibinfo {author} {L{\'e}onard, J.}\ and\
  \bibinfo {author} {Greiner, M.},\ }\bibfield  {title} {\bibinfo {title}
  {\emph {Quantum critical behaviour at the many-body localization
  transition}},\ }\href {\doibase 10.1038/s41586-019-1527-2} {\bibfield
  {journal} {\bibinfo  {journal} {Nature}\ }\textbf {\bibinfo {volume} {573}},\
  \bibinfo {pages} {385} (\bibinfo {year} {2019})}\BibitemShut {NoStop}%
\bibitem [{\citenamefont {Lee}\ \emph {et~al.}(1957)\citenamefont {Lee},
  \citenamefont {Huang},\ and\ \citenamefont {Yang}}]{lee1957}%
  \BibitemOpen
  \bibfield  {author} {\bibinfo {author} {Lee, T.~D.}, \bibinfo {author}
  {Huang, K.}\ and\ \bibinfo {author} {Yang, C.~N.},\ }\bibfield  {title}
  {\bibinfo {title} {\emph {Eigenvalues and Eigenfunctions of a Bose System of
  Hard Spheres and Its Low-Temperature Properties}},\ }\href {\doibase
  10.1103/PhysRev.106.1135} {\bibfield  {journal} {\bibinfo  {journal} {Phys.
  Rev.}\ }\textbf {\bibinfo {volume} {106}},\ \bibinfo {pages} {1135} (\bibinfo
  {year} {1957})}\BibitemShut {NoStop}%
\bibitem [{\citenamefont {Navon}\ \emph {et~al.}(2011)\citenamefont {Navon},
  \citenamefont {Piatecki}, \citenamefont {G\"unter}, \citenamefont {Rem},
  \citenamefont {Nguyen}, \citenamefont {Chevy}, \citenamefont {Krauth},\ and\
  \citenamefont {Salomon}}]{navon2011}%
  \BibitemOpen
  \bibfield  {author} {\bibinfo {author} {Navon, N.}, \bibinfo {author}
  {Piatecki, S.}, \bibinfo {author} {G\"unter, K.}, \bibinfo {author} {Rem,
  B.}, \bibinfo {author} {Nguyen, T.~C.}, \bibinfo {author} {Chevy, F.},
  \bibinfo {author} {Krauth, W.}\ and\ \bibinfo {author} {Salomon, C.},\
  }\bibfield  {title} {\bibinfo {title} {\emph {Dynamics and Thermodynamics of
  the Low-Temperature Strongly Interacting Bose Gas}},\ }\href {\doibase
  10.1103/PhysRevLett.107.135301} {\bibfield  {journal} {\bibinfo  {journal}
  {Phys. Rev. Lett.}\ }\textbf {\bibinfo {volume} {107}},\ \bibinfo {pages}
  {135301} (\bibinfo {year} {2011})}\BibitemShut {NoStop}%
\bibitem [{\citenamefont {J\o{}rgensen}\ \emph {et~al.}(2018)\citenamefont
  {J\o{}rgensen}, \citenamefont {Bruun},\ and\ \citenamefont
  {Arlt}}]{Jorgensen2018}%
  \BibitemOpen
  \bibfield  {author} {\bibinfo {author} {J\o{}rgensen, N.~B.}, \bibinfo
  {author} {Bruun, G.~M.}\ and\ \bibinfo {author} {Arlt, J.~J.},\ }\bibfield
  {title} {\bibinfo {title} {\emph {Dilute Fluid Governed by Quantum
  Fluctuations}},\ }\href {\doibase 10.1103/PhysRevLett.121.173403} {\bibfield
  {journal} {\bibinfo  {journal} {Phys. Rev. Lett.}\ }\textbf {\bibinfo
  {volume} {121}},\ \bibinfo {pages} {173403} (\bibinfo {year}
  {2018})}\BibitemShut {NoStop}%
\bibitem [{\citenamefont {Pitaevskii}\ and\ \citenamefont
  {Stringari}(1991)}]{pitaevskii1991}%
  \BibitemOpen
  \bibfield  {author} {\bibinfo {author} {Pitaevskii, L.}\ and\ \bibinfo
  {author} {Stringari, S.},\ }\bibfield  {title} {\bibinfo {title} {\emph
  {Uncertainty principle, quantum fluctuations, and broken symmetries}},\
  }\href {\doibase 10.1007/BF00682193} {\bibfield  {journal} {\bibinfo
  {journal} {Journal of Low Temperature Physics}\ }\textbf {\bibinfo {volume}
  {85}},\ \bibinfo {pages} {377} (\bibinfo {year} {1991})}\BibitemShut
  {NoStop}%
\bibitem [{\citenamefont {Lopes}\ \emph {et~al.}(2017)\citenamefont {Lopes},
  \citenamefont {Eigen}, \citenamefont {Navon}, \citenamefont {Cl\'ement},
  \citenamefont {Smith},\ and\ \citenamefont {Hadzibabic}}]{lopes2017}%
  \BibitemOpen
  \bibfield  {author} {\bibinfo {author} {Lopes, R.}, \bibinfo {author} {Eigen,
  C.}, \bibinfo {author} {Navon, N.}, \bibinfo {author} {Cl\'ement, D.},
  \bibinfo {author} {Smith, R.~P.}\ and\ \bibinfo {author} {Hadzibabic, Z.},\
  }\bibfield  {title} {\bibinfo {title} {\emph {Quantum Depletion of a
  Homogeneous Bose-Einstein Condensate}},\ }\href {\doibase
  10.1103/PhysRevLett.119.190404} {\bibfield  {journal} {\bibinfo  {journal}
  {Phys. Rev. Lett.}\ }\textbf {\bibinfo {volume} {119}},\ \bibinfo {pages}
  {190404} (\bibinfo {year} {2017})}\BibitemShut {NoStop}%
\bibitem [{\citenamefont {Ceperley}(1995)}]{Ceperley1995}%
  \BibitemOpen
  \bibfield  {author} {\bibinfo {author} {Ceperley, D.~M.},\ }\bibfield
  {title} {\bibinfo {title} {\emph {Path integrals in the theory of condensed
  helium}},\ }\href {\doibase 10.1103/RevModPhys.67.279} {\bibfield  {journal}
  {\bibinfo  {journal} {Rev. Mod. Phys.}\ }\textbf {\bibinfo {volume} {67}},\
  \bibinfo {pages} {279} (\bibinfo {year} {1995})}\BibitemShut {NoStop}%
\bibitem [{\citenamefont {Svistunov}\ \emph {et~al.}(2015)\citenamefont
  {Svistunov}, \citenamefont {Babaev},\ and\ \citenamefont
  {Prokof'ev}}]{Svistunovbook}%
  \BibitemOpen
  \bibfield  {author} {\bibinfo {author} {Svistunov, B.~V.}, \bibinfo {author}
  {Babaev, E.~S.}\ and\ \bibinfo {author} {Prokof'ev, N.~V.},\ }\href@noop {}
  {\emph {\bibinfo {title} {{Superfluid States of Matter}}}}\ (\bibinfo
  {publisher} {CRC Press},\ \bibinfo {year} {2015})\BibitemShut {NoStop}%
\bibitem [{\citenamefont {Fisher}\ \emph {et~al.}(1989)\citenamefont {Fisher},
  \citenamefont {Weichman}, \citenamefont {Grinstein},\ and\ \citenamefont
  {Fisher}}]{fisher1989}%
  \BibitemOpen
  \bibfield  {author} {\bibinfo {author} {Fisher, M. P.~A.}, \bibinfo {author}
  {Weichman, P.~B.}, \bibinfo {author} {Grinstein, G.}\ and\ \bibinfo {author}
  {Fisher, D.~S.},\ }\bibfield  {title} {\bibinfo {title} {\emph {Boson
  localization and the superfluid-insulator transition}},\ }\href {\doibase
  10.1103/PhysRevB.40.546} {\bibfield  {journal} {\bibinfo  {journal} {Phys.
  Rev. B}\ }\textbf {\bibinfo {volume} {40}},\ \bibinfo {pages} {546} (\bibinfo
  {year} {1989})}\BibitemShut {NoStop}%
\bibitem [{\citenamefont {Bloch}\ \emph {et~al.}(2008)\citenamefont {Bloch},
  \citenamefont {Dalibard},\ and\ \citenamefont {Zwerger}}]{bloch2008}%
  \BibitemOpen
  \bibfield  {author} {\bibinfo {author} {Bloch, I.}, \bibinfo {author}
  {Dalibard, J.}\ and\ \bibinfo {author} {Zwerger, W.},\ }\bibfield  {title}
  {\bibinfo {title} {\emph {Many-body physics with ultracold gases}},\ }\href
  {\doibase 10.1103/RevModPhys.80.885} {\bibfield  {journal} {\bibinfo
  {journal} {Rev. Mod. Phys.}\ }\textbf {\bibinfo {volume} {80}},\ \bibinfo
  {pages} {885} (\bibinfo {year} {2008})}\BibitemShut {NoStop}%
\bibitem [{\citenamefont {Georgescu}\ \emph {et~al.}(2014)\citenamefont
  {Georgescu}, \citenamefont {Ashhab},\ and\ \citenamefont
  {Nori}}]{Georgescu2014}%
  \BibitemOpen
  \bibfield  {author} {\bibinfo {author} {Georgescu, I.~M.}, \bibinfo {author}
  {Ashhab, S.}\ and\ \bibinfo {author} {Nori, F.},\ }\bibfield  {title}
  {\bibinfo {title} {\emph {Quantum simulation}},\ }\href {\doibase
  10.1103/RevModPhys.86.153} {\bibfield  {journal} {\bibinfo  {journal} {Rev.
  Mod. Phys.}\ }\textbf {\bibinfo {volume} {86}},\ \bibinfo {pages} {153}
  (\bibinfo {year} {2014})}\BibitemShut {NoStop}%
\bibitem [{\citenamefont {Schellekens}\ \emph {et~al.}(2005)\citenamefont
  {Schellekens}, \citenamefont {Hoppeler}, \citenamefont {Perrin},
  \citenamefont {Gomes}, \citenamefont {Boiron}, \citenamefont {Aspect},\ and\
  \citenamefont {Westbrook}}]{schellekens2005}%
  \BibitemOpen
  \bibfield  {author} {\bibinfo {author} {Schellekens, M.}, \bibinfo {author}
  {Hoppeler, R.}, \bibinfo {author} {Perrin, A.}, \bibinfo {author} {Gomes,
  J.~V.}, \bibinfo {author} {Boiron, D.}, \bibinfo {author} {Aspect, A.}\ and\
  \bibinfo {author} {Westbrook, C.~I.},\ }\bibfield  {title} {\bibinfo {title}
  {\emph {Hanbury Brown Twiss Effect for Ultracold Quantum Gases}},\ }\href
  {\doibase 10.1126/science.1118024} {\bibfield  {journal} {\bibinfo  {journal}
  {Science}\ }\textbf {\bibinfo {volume} {310}},\ \bibinfo {pages} {648}
  (\bibinfo {year} {2005})},\ \Eprint
  {http://arxiv.org/abs/https://www.science.org/doi/pdf/10.1126/science.1118024}
  {https://www.science.org/doi/pdf/10.1126/science.1118024} \BibitemShut
  {NoStop}%
\bibitem [{\citenamefont {Vassen}\ \emph {et~al.}(2012)\citenamefont {Vassen},
  \citenamefont {Cohen-Tannoudji}, \citenamefont {Leduc}, \citenamefont
  {Boiron}, \citenamefont {Westbrook}, \citenamefont {Truscott}, \citenamefont
  {Baldwin}, \citenamefont {Birkl}, \citenamefont {Cancio},\ and\ \citenamefont
  {Trippenbach}}]{vassen2012}%
  \BibitemOpen
  \bibfield  {author} {\bibinfo {author} {Vassen, W.}, \bibinfo {author}
  {Cohen-Tannoudji, C.}, \bibinfo {author} {Leduc, M.}, \bibinfo {author}
  {Boiron, D.}, \bibinfo {author} {Westbrook, C.~I.}, \bibinfo {author}
  {Truscott, A.}, \bibinfo {author} {Baldwin, K.}, \bibinfo {author} {Birkl,
  G.}, \bibinfo {author} {Cancio, P.}\ and\ \bibinfo {author} {Trippenbach,
  M.},\ }\bibfield  {title} {\bibinfo {title} {\emph {Cold and trapped
  metastable noble gases}},\ }\href {\doibase 10.1103/RevModPhys.84.175}
  {\bibfield  {journal} {\bibinfo  {journal} {Rev. Mod. Phys.}\ }\textbf
  {\bibinfo {volume} {84}},\ \bibinfo {pages} {175} (\bibinfo {year}
  {2012})}\BibitemShut {NoStop}%
\bibitem [{\citenamefont {Cayla}\ \emph {et~al.}(2018)\citenamefont {Cayla},
  \citenamefont {Carcy}, \citenamefont {Bouton}, \citenamefont {Chang},
  \citenamefont {Carleo}, \citenamefont {Mancini},\ and\ \citenamefont
  {Cl\'ement}}]{cayla2018}%
  \BibitemOpen
  \bibfield  {author} {\bibinfo {author} {Cayla, H.}, \bibinfo {author} {Carcy,
  C.}, \bibinfo {author} {Bouton, Q.}, \bibinfo {author} {Chang, R.}, \bibinfo
  {author} {Carleo, G.}, \bibinfo {author} {Mancini, M.}\ and\ \bibinfo
  {author} {Cl\'ement, D.},\ }\bibfield  {title} {\bibinfo {title} {\emph
  {Single-atom-resolved probing of lattice gases in momentum space}},\ }\href
  {\doibase 10.1103/PhysRevA.97.061609} {\bibfield  {journal} {\bibinfo
  {journal} {Phys. Rev. A}\ }\textbf {\bibinfo {volume} {97}},\ \bibinfo
  {pages} {061609} (\bibinfo {year} {2018})}\BibitemShut {NoStop}%
\bibitem [{\citenamefont {Dall}\ \emph {et~al.}(2013)\citenamefont {Dall},
  \citenamefont {Manning}, \citenamefont {Hodgman}, \citenamefont {RuGway},
  \citenamefont {Kheruntsyan},\ and\ \citenamefont {Truscott}}]{dall2013}%
  \BibitemOpen
  \bibfield  {author} {\bibinfo {author} {Dall, R.~G.}, \bibinfo {author}
  {Manning, A.~G.}, \bibinfo {author} {Hodgman, S.~S.}, \bibinfo {author}
  {RuGway, W.}, \bibinfo {author} {Kheruntsyan, K.~V.}\ and\ \bibinfo {author}
  {Truscott, A.~G.},\ }\bibfield  {title} {\bibinfo {title} {\emph {Ideal
  n-body correlations with massive particles}},\ }\href {\doibase
  10.1038/nphys2632} {\bibfield  {journal} {\bibinfo  {journal} {Nature
  Physics}\ }\textbf {\bibinfo {volume} {9}},\ \bibinfo {pages} {341} (\bibinfo
  {year} {2013})}\BibitemShut {NoStop}%
\bibitem [{\citenamefont {Herc\'e}\ \emph {et~al.}(2023)\citenamefont
  {Herc\'e}, \citenamefont {Bureik}, \citenamefont {T\'enart}, \citenamefont
  {Aspect}, \citenamefont {Dareau},\ and\ \citenamefont
  {Cl\'ement}}]{herce2023}%
  \BibitemOpen
  \bibfield  {author} {\bibinfo {author} {Herc\'e, G.}, \bibinfo {author}
  {Bureik, J.-P.}, \bibinfo {author} {T\'enart, A.}, \bibinfo {author} {Aspect,
  A.}, \bibinfo {author} {Dareau, A.}\ and\ \bibinfo {author} {Cl\'ement, D.},\
  }\bibfield  {title} {\bibinfo {title} {\emph {Full counting statistics of
  interacting lattice gases after an expansion: The role of condensate
  depletion in many-body coherence}},\ }\href {\doibase
  10.1103/PhysRevResearch.5.L012037} {\bibfield  {journal} {\bibinfo  {journal}
  {Phys. Rev. Res.}\ }\textbf {\bibinfo {volume} {5}},\ \bibinfo {pages}
  {L012037} (\bibinfo {year} {2023})}\BibitemShut {NoStop}%
\bibitem [{\citenamefont {Fetter}\ and\ \citenamefont
  {Walecka}(2012)}]{FetterWalecka}%
  \BibitemOpen
  \bibfield  {author} {\bibinfo {author} {Fetter, A.~L.}\ and\ \bibinfo
  {author} {Walecka, J.~D.},\ }\href@noop {} {\emph {\bibinfo {title} {Quantum
  Theory of Many-Particle Systems}}}\ (\bibinfo  {publisher} {Dover},\ \bibinfo
  {year} {2012})\BibitemShut {NoStop}%
\bibitem [{\citenamefont {Carlen}\ \emph {et~al.}(2021)\citenamefont {Carlen},
  \citenamefont {Holzmann}, \citenamefont {Jauslin},\ and\ \citenamefont
  {Lieb}}]{carlen2021}%
  \BibitemOpen
  \bibfield  {author} {\bibinfo {author} {Carlen, E.~A.}, \bibinfo {author}
  {Holzmann, M.}, \bibinfo {author} {Jauslin, I.}\ and\ \bibinfo {author}
  {Lieb, E.~H.},\ }\bibfield  {title} {\bibinfo {title} {\emph {Simplified
  approach to the repulsive Bose gas from low to high densities and its
  numerical accuracy}},\ }\href {\doibase 10.1103/PhysRevA.103.053309}
  {\bibfield  {journal} {\bibinfo  {journal} {Phys. Rev. A}\ }\textbf {\bibinfo
  {volume} {103}},\ \bibinfo {pages} {053309} (\bibinfo {year}
  {2021})}\BibitemShut {NoStop}%
\bibitem [{\citenamefont {Ursell}(1927)}]{ursell1927}%
  \BibitemOpen
  \bibfield  {author} {\bibinfo {author} {Ursell, H.~D.},\ }\bibfield  {title}
  {\bibinfo {title} {\emph {The evaluation of Gibbs' phase-integral for
  imperfect gases}},\ }\href {\doibase 10.1017/S0305004100011191} {\bibfield
  {journal} {\bibinfo  {journal} {Mathematical Proceedings of the Cambridge
  Philosophical Society}\ }\textbf {\bibinfo {volume} {23}},\ \bibinfo {pages}
  {685–697} (\bibinfo {year} {1927})}\BibitemShut {NoStop}%
\bibitem [{\citenamefont {Bouton}\ \emph {et~al.}(2015)\citenamefont {Bouton},
  \citenamefont {Chang}, \citenamefont {Hoendervanger}, \citenamefont
  {Nogrette}, \citenamefont {Aspect}, \citenamefont {Westbrook},\ and\
  \citenamefont {Cl\'ement}}]{bouton2015}%
  \BibitemOpen
  \bibfield  {author} {\bibinfo {author} {Bouton, Q.}, \bibinfo {author}
  {Chang, R.}, \bibinfo {author} {Hoendervanger, A.~L.}, \bibinfo {author}
  {Nogrette, F.}, \bibinfo {author} {Aspect, A.}, \bibinfo {author} {Westbrook,
  C.~I.}\ and\ \bibinfo {author} {Cl\'ement, D.},\ }\bibfield  {title}
  {\bibinfo {title} {\emph {Fast production of Bose-Einstein condensates of
  metastable helium}},\ }\href {\doibase 10.1103/PhysRevA.91.061402} {\bibfield
   {journal} {\bibinfo  {journal} {Phys. Rev. A}\ }\textbf {\bibinfo {volume}
  {91}},\ \bibinfo {pages} {061402} (\bibinfo {year} {2015})}\BibitemShut
  {NoStop}%
\bibitem [{\citenamefont {Carcy}\ \emph {et~al.}(2021)\citenamefont {Carcy},
  \citenamefont {Herc\'e}, \citenamefont {Tenart}, \citenamefont {Roscilde},\
  and\ \citenamefont {Cl\'ement}}]{carcy2021}%
  \BibitemOpen
  \bibfield  {author} {\bibinfo {author} {Carcy, C.}, \bibinfo {author}
  {Herc\'e, G.}, \bibinfo {author} {Tenart, A.}, \bibinfo {author} {Roscilde,
  T.}\ and\ \bibinfo {author} {Cl\'ement, D.},\ }\bibfield  {title} {\bibinfo
  {title} {\emph {Certifying the Adiabatic Preparation of Ultracold Lattice
  Bosons in the Vicinity of the Mott Transition}},\ }\href {\doibase
  10.1103/PhysRevLett.126.045301} {\bibfield  {journal} {\bibinfo  {journal}
  {Phys. Rev. Lett.}\ }\textbf {\bibinfo {volume} {126}},\ \bibinfo {pages}
  {045301} (\bibinfo {year} {2021})}\BibitemShut {NoStop}%
\bibitem [{\citenamefont {Nogrette}\ \emph {et~al.}(2015)\citenamefont
  {Nogrette}, \citenamefont {Heurteau}, \citenamefont {Chang}, \citenamefont
  {Bouton}, \citenamefont {Westbrook}, \citenamefont {Sellem},\ and\
  \citenamefont {Clément}}]{nogrette2015}%
  \BibitemOpen
  \bibfield  {author} {\bibinfo {author} {Nogrette, F.}, \bibinfo {author}
  {Heurteau, D.}, \bibinfo {author} {Chang, R.}, \bibinfo {author} {Bouton,
  Q.}, \bibinfo {author} {Westbrook, C.~I.}, \bibinfo {author} {Sellem, R.}\
  and\ \bibinfo {author} {Clément, D.},\ }\bibfield  {title} {\bibinfo {title}
  {\emph {Characterization of a detector chain using a FPGA-based
  time-to-digital converter to reconstruct the three-dimensional coordinates of
  single particles at high flux}},\ }\href {\doibase 10.1063/1.4935474}
  {\bibfield  {journal} {\bibinfo  {journal} {Review of Scientific
  Instruments}\ }\textbf {\bibinfo {volume} {86}},\ \bibinfo {pages} {113105}
  (\bibinfo {year} {2015})}\BibitemShut {NoStop}%
\bibitem [{\citenamefont {Tenart}\ \emph {et~al.}(2020)\citenamefont {Tenart},
  \citenamefont {Carcy}, \citenamefont {Cayla}, \citenamefont {Bourdel},
  \citenamefont {Mancini},\ and\ \citenamefont {Cl\'ement}}]{tenart2020}%
  \BibitemOpen
  \bibfield  {author} {\bibinfo {author} {Tenart, A.}, \bibinfo {author}
  {Carcy, C.}, \bibinfo {author} {Cayla, H.}, \bibinfo {author} {Bourdel, T.},
  \bibinfo {author} {Mancini, M.}\ and\ \bibinfo {author} {Cl\'ement, D.},\
  }\bibfield  {title} {\bibinfo {title} {\emph {Two-body collisions in the
  time-of-flight dynamics of lattice Bose superfluids}},\ }\href {\doibase
  10.1103/PhysRevResearch.2.013017} {\bibfield  {journal} {\bibinfo  {journal}
  {Phys. Rev. Res.}\ }\textbf {\bibinfo {volume} {2}},\ \bibinfo {pages}
  {013017} (\bibinfo {year} {2020})}\BibitemShut {NoStop}%
\bibitem [{\citenamefont {Herc\'e}\ \emph {et~al.}(2021)\citenamefont
  {Herc\'e}, \citenamefont {Carcy}, \citenamefont {Tenart}, \citenamefont
  {Bureik}, \citenamefont {Dareau}, \citenamefont {Cl\'ement},\ and\
  \citenamefont {Roscilde}}]{herce2021}%
  \BibitemOpen
  \bibfield  {author} {\bibinfo {author} {Herc\'e, G.}, \bibinfo {author}
  {Carcy, C.}, \bibinfo {author} {Tenart, A.}, \bibinfo {author} {Bureik,
  J.-P.}, \bibinfo {author} {Dareau, A.}, \bibinfo {author} {Cl\'ement, D.}\
  and\ \bibinfo {author} {Roscilde, T.},\ }\bibfield  {title} {\bibinfo {title}
  {\emph {Studying the low-entropy Mott transition of bosons in a
  three-dimensional optical lattice by measuring the full momentum-space
  density}},\ }\href {\doibase 10.1103/PhysRevA.104.L011301} {\bibfield
  {journal} {\bibinfo  {journal} {Phys. Rev. A}\ }\textbf {\bibinfo {volume}
  {104}},\ \bibinfo {pages} {L011301} (\bibinfo {year} {2021})}\BibitemShut
  {NoStop}%
\bibitem [{\citenamefont {Toth}\ \emph {et~al.}(2008)\citenamefont {Toth},
  \citenamefont {Rey},\ and\ \citenamefont {Blakie}}]{toth2008}%
  \BibitemOpen
  \bibfield  {author} {\bibinfo {author} {Toth, E.}, \bibinfo {author} {Rey,
  A.~M.}\ and\ \bibinfo {author} {Blakie, P.~B.},\ }\bibfield  {title}
  {\bibinfo {title} {\emph {Theory of correlations between ultracold bosons
  released from an optical lattice}},\ }\href {\doibase
  10.1103/PhysRevA.78.013627} {\bibfield  {journal} {\bibinfo  {journal} {Phys.
  Rev. A}\ }\textbf {\bibinfo {volume} {78}},\ \bibinfo {pages} {013627}
  (\bibinfo {year} {2008})}\BibitemShut {NoStop}%
\bibitem [{\citenamefont {Cayla}\ \emph {et~al.}(2020)\citenamefont {Cayla},
  \citenamefont {Butera}, \citenamefont {Carcy}, \citenamefont {Tenart},
  \citenamefont {Herc\'e}, \citenamefont {Mancini}, \citenamefont {Aspect},
  \citenamefont {Carusotto},\ and\ \citenamefont {Cl\'ement}}]{cayla2020}%
  \BibitemOpen
  \bibfield  {author} {\bibinfo {author} {Cayla, H.}, \bibinfo {author}
  {Butera, S.}, \bibinfo {author} {Carcy, C.}, \bibinfo {author} {Tenart, A.},
  \bibinfo {author} {Herc\'e, G.}, \bibinfo {author} {Mancini, M.}, \bibinfo
  {author} {Aspect, A.}, \bibinfo {author} {Carusotto, I.}\ and\ \bibinfo
  {author} {Cl\'ement, D.},\ }\bibfield  {title} {\bibinfo {title} {\emph
  {Hanbury Brown and Twiss Bunching of Phonons and of the Quantum Depletion in
  an Interacting Bose Gas}},\ }\href {\doibase 10.1103/PhysRevLett.125.165301}
  {\bibfield  {journal} {\bibinfo  {journal} {Phys. Rev. Lett.}\ }\textbf
  {\bibinfo {volume} {125}},\ \bibinfo {pages} {165301} (\bibinfo {year}
  {2020})}\BibitemShut {NoStop}%
\bibitem [{\citenamefont {Wallin}\ \emph {et~al.}(1994)\citenamefont {Wallin},
  \citenamefont {So/rensen}, \citenamefont {Girvin},\ and\ \citenamefont
  {Young}}]{Wallinetal1994}%
  \BibitemOpen
  \bibfield  {author} {\bibinfo {author} {Wallin, M.}, \bibinfo {author}
  {So/rensen, E.~S.}, \bibinfo {author} {Girvin, S.~M.}\ and\ \bibinfo {author}
  {Young, A.~P.},\ }\bibfield  {title} {\bibinfo {title} {\emph
  {Superconductor-insulator transition in two-dimensional dirty boson
  systems}},\ }\href {\doibase 10.1103/PhysRevB.49.12115} {\bibfield  {journal}
  {\bibinfo  {journal} {Phys. Rev. B}\ }\textbf {\bibinfo {volume} {49}},\
  \bibinfo {pages} {12115} (\bibinfo {year} {1994})}\BibitemShut {NoStop}%
\bibitem [{\citenamefont {Roscilde}(2008)}]{Roscilde2008}%
  \BibitemOpen
  \bibfield  {author} {\bibinfo {author} {Roscilde, T.},\ }\bibfield  {title}
  {\bibinfo {title} {\emph {Bosons in one-dimensional incommensurate
  superlattices}},\ }\href {\doibase 10.1103/PhysRevA.77.063605} {\bibfield
  {journal} {\bibinfo  {journal} {Phys. Rev. A}\ }\textbf {\bibinfo {volume}
  {77}},\ \bibinfo {pages} {063605} (\bibinfo {year} {2008})}\BibitemShut
  {NoStop}%
\bibitem [{\citenamefont {Fang}\ \emph {et~al.}(2016)\citenamefont {Fang},
  \citenamefont {Johnson}, \citenamefont {Roscilde},\ and\ \citenamefont
  {Bouchoule}}]{Fangetal2016}%
  \BibitemOpen
  \bibfield  {author} {\bibinfo {author} {Fang, B.}, \bibinfo {author}
  {Johnson, A.}, \bibinfo {author} {Roscilde, T.}\ and\ \bibinfo {author}
  {Bouchoule, I.},\ }\bibfield  {title} {\bibinfo {title} {\emph
  {Momentum-Space Correlations of a One-Dimensional Bose Gas}},\ }\href
  {\doibase 10.1103/PhysRevLett.116.050402} {\bibfield  {journal} {\bibinfo
  {journal} {Phys. Rev. Lett.}\ }\textbf {\bibinfo {volume} {116}},\ \bibinfo
  {pages} {050402} (\bibinfo {year} {2016})}\BibitemShut {NoStop}%
\bibitem [{\citenamefont {Ferioli}\ \emph {et~al.}(2024)\citenamefont
  {Ferioli}, \citenamefont {Pancaldi}, \citenamefont {Glicenstein},
  \citenamefont {Cl\'ement}, \citenamefont {Browaeys},\ and\ \citenamefont
  {Ferrier-Barbut}}]{ferioli2024}%
  \BibitemOpen
  \bibfield  {author} {\bibinfo {author} {Ferioli, G.}, \bibinfo {author}
  {Pancaldi, S.}, \bibinfo {author} {Glicenstein, A.}, \bibinfo {author}
  {Cl\'ement, D.}, \bibinfo {author} {Browaeys, A.}\ and\ \bibinfo {author}
  {Ferrier-Barbut, I.},\ }\bibfield  {title} {\bibinfo {title} {\emph
  {Non-Gaussian Correlations in the Steady State of Driven-Dissipative Clouds
  of Two-Level Atoms}},\ }\href {\doibase 10.1103/PhysRevLett.132.133601}
  {\bibfield  {journal} {\bibinfo  {journal} {Phys. Rev. Lett.}\ }\textbf
  {\bibinfo {volume} {132}},\ \bibinfo {pages} {133601} (\bibinfo {year}
  {2024})}\BibitemShut {NoStop}%
\bibitem [{\citenamefont {Rom}\ \emph {et~al.}(2006)\citenamefont {Rom},
  \citenamefont {Best}, \citenamefont {van Oosten}, \citenamefont {Schneider},
  \citenamefont {F{\"o}lling}, \citenamefont {Paredes},\ and\ \citenamefont
  {Bloch}}]{Rom2006}%
  \BibitemOpen
  \bibfield  {author} {\bibinfo {author} {Rom, T.}, \bibinfo {author} {Best,
  T.}, \bibinfo {author} {van Oosten, D.}, \bibinfo {author} {Schneider, U.},
  \bibinfo {author} {F{\"o}lling, S.}, \bibinfo {author} {Paredes, B.}\ and\
  \bibinfo {author} {Bloch, I.},\ }\bibfield  {title} {\bibinfo {title} {\emph
  {Free fermion antibunching in a degenerate atomic Fermi gas released from an
  optical lattice}},\ }\href {\doibase 10.1038/nature05319} {\bibfield
  {journal} {\bibinfo  {journal} {Nature}\ }\textbf {\bibinfo {volume} {444}},\
  \bibinfo {pages} {733} (\bibinfo {year} {2006})}\BibitemShut {NoStop}%
\bibitem [{\citenamefont {Ténart}(2021)}]{TenartPhD}%
  \BibitemOpen
  \bibfield  {author} {\bibinfo {author} {Ténart, A.},\ }\bibfield  {title}
  {\bibinfo {title} {\emph {Momentum-space correlations in the depletion of
  weakly interacting lattice Bose gases}},\ }\href
  {http://www.theses.fr/2021UPASP128} {\  (\bibinfo {year} {2021})},\ \bibinfo
  {note} {thèse de doctorat dirigée par D. Clément, Physique université
  Paris-Saclay 2021}\BibitemShut {NoStop}%
\bibitem [{\citenamefont {Carcy}\ \emph {et~al.}(2019)\citenamefont {Carcy},
  \citenamefont {Cayla}, \citenamefont {Tenart}, \citenamefont {Aspect},
  \citenamefont {Mancini},\ and\ \citenamefont {Cl\'ement}}]{carcy2019}%
  \BibitemOpen
  \bibfield  {author} {\bibinfo {author} {Carcy, C.}, \bibinfo {author} {Cayla,
  H.}, \bibinfo {author} {Tenart, A.}, \bibinfo {author} {Aspect, A.}, \bibinfo
  {author} {Mancini, M.}\ and\ \bibinfo {author} {Cl\'ement, D.},\ }\bibfield
  {title} {\bibinfo {title} {\emph {Momentum-Space Atom Correlations in a Mott
  Insulator}},\ }\href {\doibase 10.1103/PhysRevX.9.041028} {\bibfield
  {journal} {\bibinfo  {journal} {Phys. Rev. X}\ }\textbf {\bibinfo {volume}
  {9}},\ \bibinfo {pages} {041028} (\bibinfo {year} {2019})}\BibitemShut
  {NoStop}%
\bibitem [{\citenamefont {Dainty}(1975)}]{Dainty-book}%
  \BibitemOpen
  \bibinfo {editor} {Dainty, J.~C.},\ ed.,\ \href@noop {} {\emph {\bibinfo
  {title} {Laser speckle and related phenomena}}},\ \bibinfo {edition} {1st}\
  ed.\ (\bibinfo  {publisher} {Springer},\ \bibinfo {year} {1975})\BibitemShut
  {NoStop}%
\bibitem [{\citenamefont {Sondhi}\ \emph {et~al.}(1997)\citenamefont {Sondhi},
  \citenamefont {Girvin}, \citenamefont {Carini},\ and\ \citenamefont
  {Shahar}}]{SondhiG1997}%
  \BibitemOpen
  \bibfield  {author} {\bibinfo {author} {Sondhi, S.~L.}, \bibinfo {author}
  {Girvin, S.~M.}, \bibinfo {author} {Carini, J.~P.}\ and\ \bibinfo {author}
  {Shahar, D.},\ }\bibfield  {title} {\bibinfo {title} {\emph {Continuous
  quantum phase transitions}},\ }\href {\doibase 10.1103/RevModPhys.69.315}
  {\bibfield  {journal} {\bibinfo  {journal} {Rev. Mod. Phys.}\ }\textbf
  {\bibinfo {volume} {69}},\ \bibinfo {pages} {315} (\bibinfo {year}
  {1997})}\BibitemShut {NoStop}%
\bibitem [{\citenamefont {Roscilde}\ \emph {et~al.}(2016)\citenamefont
  {Roscilde}, \citenamefont {Faulkner}, \citenamefont {Bramwell},\ and\
  \citenamefont {Holdsworth}}]{Roscildeetal2016}%
  \BibitemOpen
  \bibfield  {author} {\bibinfo {author} {Roscilde, T.}, \bibinfo {author}
  {Faulkner, M.~F.}, \bibinfo {author} {Bramwell, S.~T.}\ and\ \bibinfo
  {author} {Holdsworth, P. C.~W.},\ }\bibfield  {title} {\bibinfo {title}
  {\emph {From quantum to thermal topological-sector fluctuations of strongly
  interacting Bosons in a ring lattice}},\ }\href {\doibase
  10.1088/1367-2630/18/7/075003} {\bibfield  {journal} {\bibinfo  {journal}
  {New Journal of Physics}\ }\textbf {\bibinfo {volume} {18}},\ \bibinfo
  {pages} {075003} (\bibinfo {year} {2016})}\BibitemShut {NoStop}%
\bibitem [{\citenamefont {Trotzky}\ \emph {et~al.}(2010)\citenamefont
  {Trotzky}, \citenamefont {Pollet}, \citenamefont {Gerbier}, \citenamefont
  {Schnorrberger}, \citenamefont {Bloch}, \citenamefont {Prokof’ev},
  \citenamefont {Svistunov},\ and\ \citenamefont {Troyer}}]{trotzky2010}%
  \BibitemOpen
  \bibfield  {author} {\bibinfo {author} {Trotzky, S.}, \bibinfo {author}
  {Pollet, L.}, \bibinfo {author} {Gerbier, F.}, \bibinfo {author}
  {Schnorrberger, U.}, \bibinfo {author} {Bloch, I.}, \bibinfo {author}
  {Prokof’ev, N.~V.}, \bibinfo {author} {Svistunov, B.}\ and\ \bibinfo
  {author} {Troyer, M.},\ }\bibfield  {title} {\bibinfo {title} {\emph
  {Suppression of the critical temperature for superfluidity near the Mott
  transition}},\ }\href@noop {} {\bibfield  {journal} {\bibinfo  {journal}
  {Nature Physics}\ }\textbf {\bibinfo {volume} {6}},\ \bibinfo {pages} {998}
  (\bibinfo {year} {2010})}\BibitemShut {NoStop}%
\bibitem [{\citenamefont {Wessel}\ \emph {et~al.}(2004)\citenamefont {Wessel},
  \citenamefont {Alet}, \citenamefont {Troyer},\ and\ \citenamefont
  {Batrouni}}]{SSE}%
  \BibitemOpen
  \bibfield  {author} {\bibinfo {author} {Wessel, S.}, \bibinfo {author} {Alet,
  F.}, \bibinfo {author} {Troyer, M.}\ and\ \bibinfo {author} {Batrouni,
  G.~G.},\ }\bibfield  {title} {\bibinfo {title} {\emph {Quantum Monte Carlo
  simulations of confined bosonic atoms in optical lattices}},\ }\href
  {\doibase 10.1103/PhysRevA.70.053615} {\bibfield  {journal} {\bibinfo
  {journal} {Phys. Rev. A}\ }\textbf {\bibinfo {volume} {70}},\ \bibinfo
  {pages} {053615} (\bibinfo {year} {2004})}\BibitemShut {NoStop}%
\end{thebibliography}%
%%%%%%%%%%%%%%%%%%%%%%%%%%%%%%%%%%%%%%%

\cleardoublepage
\begin{widetext}
\begin{center}
\textbf{\large SUPPLEMENTAL MATERIAL – Suppression of Bogoliubov momentum pairing and emergence of non-Gaussian correlations in ultracold interacting Bose gases}
\\
\vspace{5mm}
Jan-Philipp Bureik, Ga\'etan Herc\'e, Maxime Allemand, Antoine Tenart, Tommaso Roscilde, David Cl\'ement
\vspace{5mm}
\end{center}
\end{widetext}

\setcounter{section}{0}
\setcounter{figure}{0} 
\setcounter{equation}{0} 
\renewcommand\theequation{S\arabic{equation}} 
\renewcommand\thefigure{S\arabic{figure}}  

%%%%%%%%%%%%%%%%%%%%%%%%%%%%%%%%%%%%%%%

\section{Thermometry of the lattice Bose gases}

To estimate the temperature of the lattice Bose gases realised in the experiment, we compare the measured momentum densities $\rho({\bm k})$ to ab-initio calculations for the Bose-Hubbard hamiltonian (BHH). More specifically, we perform unbiased Quantum Monte-Carlo calculations of the momentum density $\rho_{\rm QMC}({\bm k},T)$ for the BHH model using all the calibrated experimental parameters (particle number, trapping potentials), except for the temperature $T$ which is not measured in the experiment. At each value of $U/J$, we minimise the distance between $\rho({\bm k})$ and $\rho_{\rm QMC}({\bm k},T)$ to estimate the temperature $T$ in the experiment. This procedure is described in \cite{carcy2021}.

For the data shown in this work, the estimated reduced temperature varies smoothly from $T/J=0.8(3)$ at $U/J=1$ to $T/J=3.1(2)$ at $U/J=22$. This increase is compatible with an isentropic variation of $U/J$ \cite{carcy2021}, as stated in the main text.

\section{Two-body connected correlations in the experiment}

From the measured 3D distributions of individual atoms in momentum space, we calculate the connected two-body correlations $G_{c}^{(2)}({\bm k}, {\bm k}')=\langle N({\bm k}) N({\bm k}')\rangle - \langle N({\bm k}) \rangle \langle N({\bm k}')\rangle$. More specifically, we evaluate the terms  $\langle N({\bm k}) N({\bm k}')\rangle $ and $\langle N({\bm k}) \rangle \langle N({\bm k}')\rangle$ independently in the depletion of interacting Bose gases (in the volume $\Omega_{k}$ shown in Fig.~2a of the main text) proceeding similarly to the method introduced in \cite{tenart2021}. 
In short, we first consider all the couples of detected atoms $({\bm k},{\bm k}')$ in a given shot and fill a 3D histogram with the sum of their momenta, ${\bm K}={\bm k}+{\bm k}'$. Here the elementary volume of the histogram is a cube of size $0.012 k_d$. The term $\langle N({\bm k}) N({\bm k}')\rangle$ is obtained from averaging the 3D histogram over all the shots. The second term, $\langle N({\bm k}) \rangle \langle N({\bm k}')\rangle$, is obtained from first merging the atoms from all shots together and subsequently computing the 3D histograms of their sum ${\bm K}$.

Atoms with opposite momenta populate the ${\bm K}\simeq {\bm 0}$ region of the histogram. Considering an elementary volume located at a distance $\delta {\bm k}$ from ${\bm K}={\bm 0}$ provides us with the value of $G^{(2)}_c({\bm K} = \delta \bm{k}) = \sum_{\bm k} G^{(2)}_c({\bm k},-{\bm k}+ \delta \bm{k})$. In addition, we eliminate an artifact due to some ambiguity in the reconstruction of the 3D momenta of the atoms from the timestamps of the raw detection events. This artifact occurs at low values of $U/J$ for which the density of the condensate peak located at ${\bm k}={\bm 0}$ is high enough to induce this effect that would indeed contribute to the connected correlations at opposite momenta (see \cite{TenartPhD} for further details).

The two-body connected correlation $\bar{G}^{(2)}_c(\mathbf{0})$ is then obtained from integrating $G^{(2)}_c(\bm{K}=\delta \bm{k})$ over a cubic volume of size $\Delta k$:
\begin{equation*}
    \begin{split}
        \bar{G}^{(2)}_c(\bm{0})  
        = \int_{[-\Delta k/2,\Delta k/2]^3} d\delta \bm{k} \ G^{(2)}_c(\bm{K}=\delta \bm{k}). 
    \end{split}
\end{equation*}
In the experiment, the normalised correlation function $g^{(2)}(\bm k, -\bm k + \delta \bm{k})$ is expected to be isotropic due to the presence of an isotropic harmonic trap (of frequency $\omega/2\pi=140(5) \sqrt{s}~$Hz where $s$ the lattice amplitude in recoil energy). In the regime of moderate interactions where a peak is indeed observed in $g^{(2)}(\bm k, -\bm k + \delta \bm{k})$, this peak is empirically found to be well fitted by an isotropic Gaussian function. We introduce the correlation length $l_c$ as the size of $g^{(2)}(\bm k, -\bm k + \delta \bm{k})$ at $1/e^2$:
\begin{equation}
\label{eq: g2}
    g^{(2)}(\bm k, -\bm k + \delta \bm{k}) = 1 + \frac{A \sqrt{2}}{\sqrt{\pi}l_c} e^{-\frac{2 \delta \bm{k} ^2}{l_c^2}}.
\end{equation}
Thus 95\% of atom pairs are expected to be found in the spherical volume where $|\delta \bm{k}| \in [0, l_c]$.
Since $G^{(2)}_c(\bm{K}=\delta \bm{k}) \propto \sum_{\bm k} \ g^{(2)}(\bm k, -\bm k + \delta \bm{k}) - 1$ it follows from Eq.~(\ref{eq: g2})  that
\begin{equation}
\label{eq: g2_int}
    \bar{G}^{(2)}_c(\bm{0}) \propto \left[\mathrm{erf}\left( \frac{\Delta k}{\sqrt{2}l_c} \right)\right]^3.
\end{equation}

\begin{figure}[h!]
  \centering
  \includegraphics[width=\linewidth]{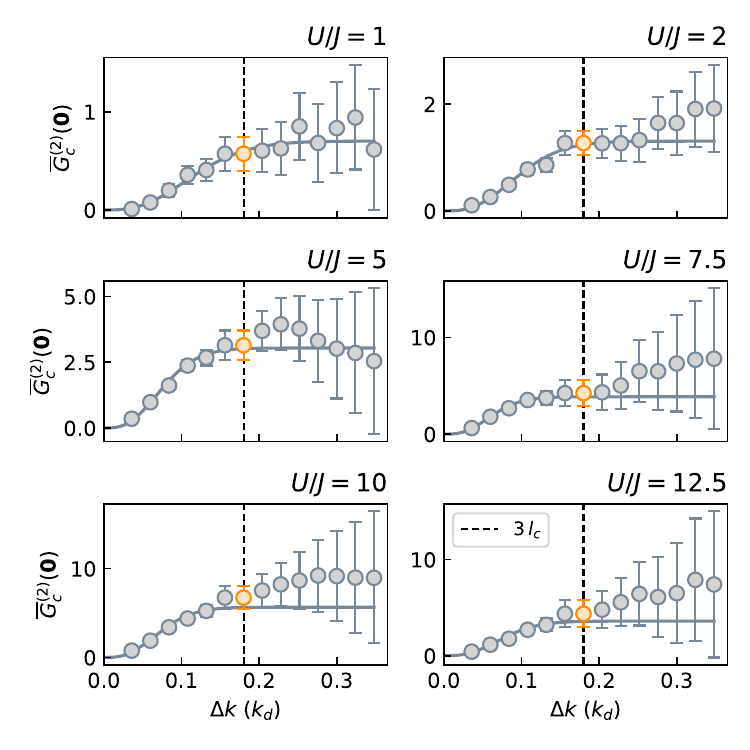}
  \caption{Two-body connected correlation $\bar{G}^{(2)}_c(\bm 0)$ as a function of the size of the integration volume $\Delta k$, from $U/J = 1$ through $12.5$. The error bars correspond to 68\%-confidence intervals obtained from a bootstrap method. The grey solid line is the empirical model following Eq.~(\ref{eq: g2_int}) associated with the correlation length $l_{c}$ of the normalised two-body correlation $g^{(2)}(\bm k, -\bm k + \delta \bm{k})$. The orange data correspond to an integration of $\Delta k = 0.18\ k_d$ and are the ones shown in the main text. The black dashed lines indicate the integration corresponding to three times the correlation length $l_c$.}
  \label{fig-sup2}
\end{figure}

At moderate interaction strengths $U/J<20$, this is verified in the experiment, as illustrated in Fig.~\ref{fig-sup2} where we plot $\bar{G}^{(2)}_c(\bm 0)$ as a function of $\Delta k$. We find indeed that $\bar{G}^{(2)}_c(\bm 0)$ increases and saturates at the expected range $\Delta k \simeq 3 \times l_c  \sim 0.18\ k_d$, with the variation $\left[\mathrm{erf}\left( \frac{\Delta k}{\sqrt{2}l_c} \right)\right]^3$ obtained from eq.~(\ref{eq: g2_int}). This is reminiscent of the fact that we observe a peak in the normalised correlation $g^{(2)}(\bm k, -\bm k + \delta \bm{k})$ centered on $\delta \bm{k}={\bm 0}$ for $U/J < 20$. To plot the amplitudes of connected correlation in the main text, we choose the volume $\Delta k^3$ such that the above integral on $d \delta {\bm k}$ saturates: we set $\Delta k = 0.18\ k_d $ such that $\bar{G}^{(2)}_c(\bm{0})$ accounts for 99\% of the pairing amplitude, theoretically. At higher $\Delta k$ only uncorrelated atoms are being included in $\bar{G}^{(2)}_c(\bm{0})$ and thus its value barely changes within error bars. However, since uncorrelated atoms contribute (equally) to the two terms in $G_{c}^{(2)}({\bm k}, {\bm k}')$, the error bars increase as a consequence of $G_{c}^{(2)}({\bm k}, {\bm k}')$ being a (constant) difference of increasingly large numbers.

\begin{figure}[h!]
  \centering
  \includegraphics[width=\linewidth]{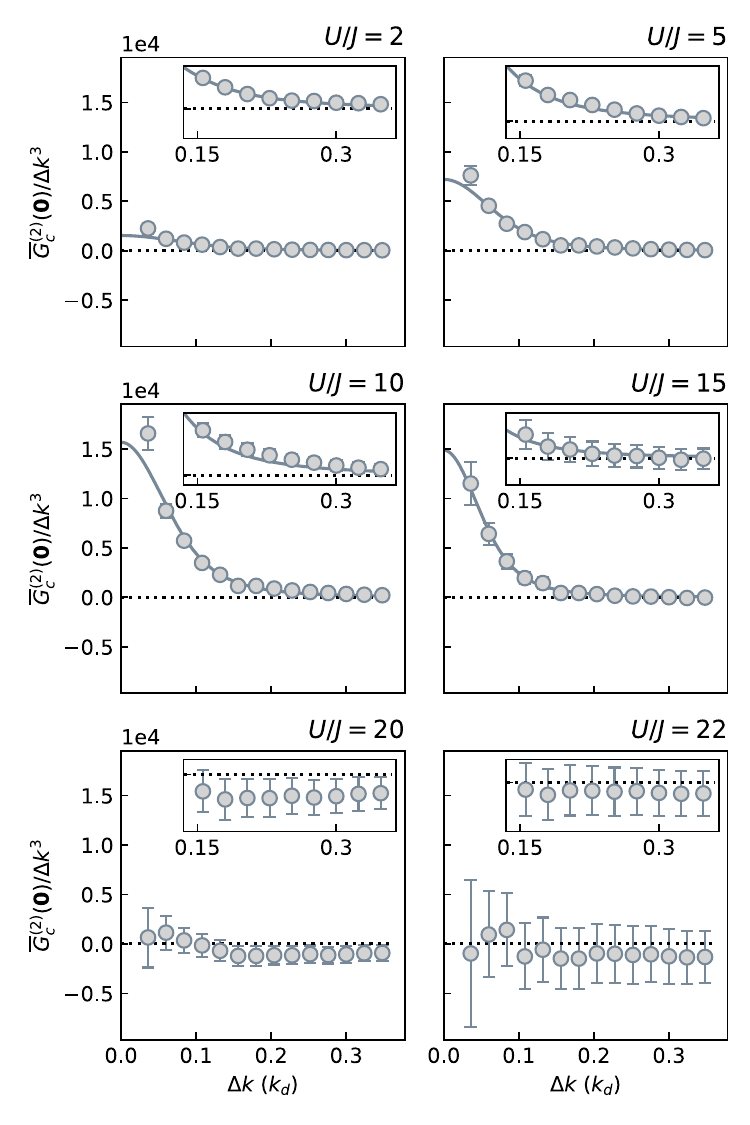}
  \caption{Normalised two-body connected correlation $\bar{G}^{(2)}_c(\bm 0)/\Delta k^3$ as a function of the size of the integration volume $\Delta k$, at $U/J = 2,\ 5,\ 10,\ 15,\ 20\ \mathrm{and}\ 22$. The error bars correspond to 68\%-confidence intervals obtained from a bootstrap method. The orange solid line is the empirical model following Eq.~(\ref{eq: g2_int}) normalised by the integration volume $\Delta k^3$. The insets show a zoom in the range $d \delta {\bm k} \in [0.14~k_{d}, 0.35~k_{d}]$. }
  \label{fig-sup3}
\end{figure}

A complementary visualisation consists in plotting an intensive quantity: $\bar{G}^{(2)}_c(\bm 0)$ normalised by the integration volume $ \Delta k^3$ as shown in Fig.~\ref{fig-sup3}. We find that $\bar{G}^{(2)}_c(\bm 0)/\Delta k^3$ decays to zero over the expected range $\Delta k \simeq 3 \times l_c  \sim 0.18\ k_d$ with the expected variation.

In contrast, at $U/J=20$ (and similarly at $U/J=22$), $\bar{G}^{(2)}_c(\bm 0)/\Delta k^3$ follows a very different trend. While it may take slightly positive values at vanishingly small values of $\Delta k$ (where the amplitude is nonetheless compatible with being equal to zero), it is constant and negative otherwise. This different behaviour is at the origin of the negative values $\bar{G}^{(2)}_c(\bm 0)$ plotted in Fig.~2 of the main text. Observing  $\bar{G}^{(2)}_c(\bm 0)<0$ implies that the number of coincidence counts in modes at opposite momenta is smaller than the coincidence counts expected in an uncorrelated ensemble of atoms with a similar momentum density $\langle N({\bm k}) \rangle=\langle a^{\dagger}_{\bm{k}} a_{\bm{k}}  \rangle $, and hence that modes at opposite momenta are anti-correlated. 

We interpret the fact that $\bar{G}^{(2)}_c(\bm 0) / \Delta k^3$ is negative and constant for most values of $\Delta k$ (see Fig.~\ref{fig-sup3} panel $U/J=20$) as an indication that the observed anti-correlation is not due to correlations between pairs of modes possessing a characteristic decay length. Indeed, two-mode correlations are generally expected to vanish beyond a given distance (in momentum space). A celebrated example of such a behavior is provided by the anti-bunching of ideal fermions \cite{Rom2006} which is present only at short distances. 
Instead when the anti-correlation originates from correlations between three bodies (or more), two-body correlation functions may not vary with the two-body distance.

\section{Normalised two-body correlations in the experiment}

In the main text, we focus on connected correlations at opposite momenta, $G_{c}^{(2)}({\bm k}, -{\bm k}+\delta {\bm k})=\langle N({\bm k}) N({-\bm k} + \delta {\bm k})\rangle - \langle N({\bm k}) \rangle \langle N({-\bm k} + \delta {\bm k})\rangle$. Alternatively, we could have used a normalised form of correlations defined as 
\begin{equation}
g^{(2)}(\bm k,\bm k')= \frac{ \langle N({\bm k}) N({\bm k}')\rangle }{ \langle N({\bm k}) \rangle \langle N({\bm k}')\rangle}.
\end{equation}
In this section, we present our measurements of the normalised two-body correlation function $g^{(2)}(\bm k,\bm k')$ and we discuss the motivations to use non-normalised connected correlations $G_{c}^{(2)}(\delta {\bm k})$ to study the pairing strength.

In the weakly-interacting regime, at small values of $U/J$, the two-body correlation function $g^{(2)}(\bm k,\bm k')$ exhibits two peaks \cite{tenart2021}: one located at $\bm k' = \bm k$ due to bosonic bunching (also referred to as Hanbury-Brown and Twiss (HBT) correlations) and one located at  $\bm k' = - \bm k$ due to interaction-induced quantum fluctuations. We are interested in extracting both the amplitudes of the bunching and of the pairing.

\begin{figure}[h!]
  \centering
  \includegraphics[width=\linewidth]{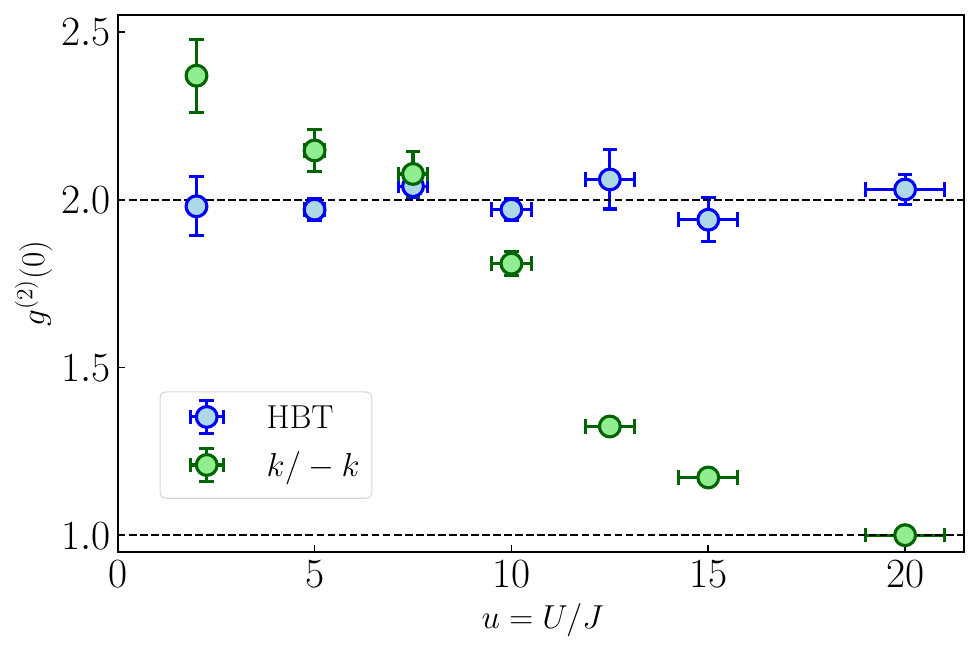}
  \caption{Magnitudes $g_{\rm HBT}^{(2)}(0)$ of local $k/k$ two-body correlations (bunching) and $g_{\rm k/-k}^{(2)}(0)$ of normalised two-body correlations at opposite momenta $k/-k$ plotted as a function of the interaction strength $u$ for data sets shown in Fig.~2 of the main text. Vertical error bars correspond to one standard deviation.}
  \label{fig-sup1}
\end{figure}

Proceeding as in \cite{tenart2021} and along the lines described above in the case of $G_{c}^{(2)}({\bm k}, {\bm k}')$, we compute the terms $\langle N({\bm k}) N({\bm k}')\rangle$ and $\langle N({\bm k}) \rangle \langle N({\bm k}')\rangle$. We then obtain $g^{(2)}(\bm k,\bm k')$ from taking their ratio. From studying the variations of the normalised correlation functions with the volume used to compute them \cite{tenart2021}, we extrapolate the amplitudes, $g_{\rm HBT}^{(2)}(0)$ and $g_{\rm k/-k}^{(2)}(0)$, at zero transverse integration, with the results plotted in Fig.~\ref{fig-sup1}.

The bunching amplitude is perfectly-contrasted in all regimes of interaction, $g_{\rm HBT}^{(2)}(0) = 2.00(5)$, extending previous observations \cite{cayla2020} to the strongly-interacting regime. In contrast, $g_{k/-k}^{(2)}(0)$ is found to decrease with the interaction $u$. This decrease is compatible with the one expected in Bogoliubov theory: the magnitude $g_{k/-k}^{(2)}(0)-1$ is indeed expected to decrease with the inverse of the mode population $1/\langle N(\bm k) \rangle$, a property of two-mode squeezed states \cite{tenart2021}. 

Note that no qualitative change is expected when the gas enters the strongly correlated regime where the Bogoliubov approximation fails: $g_{k/-k}^{(2)}(0)$ also decreases when the population of two-mode squeezed states decreases. This illustrates that normalised correlation functions are not suited to study the hierarchy of momentum-correlated subsets introduced in the main text (see Fig.~1a) and it motivates the use of connected correlations to reveal the correlated subsets of $n$ modes.

\section{Scaling of the uncertainty on the value of $G^{(2)}_{c}({\bf 0})$ with interactions}

\begin{figure}[ht!]
  \centering
  \includegraphics[width=\linewidth]{ 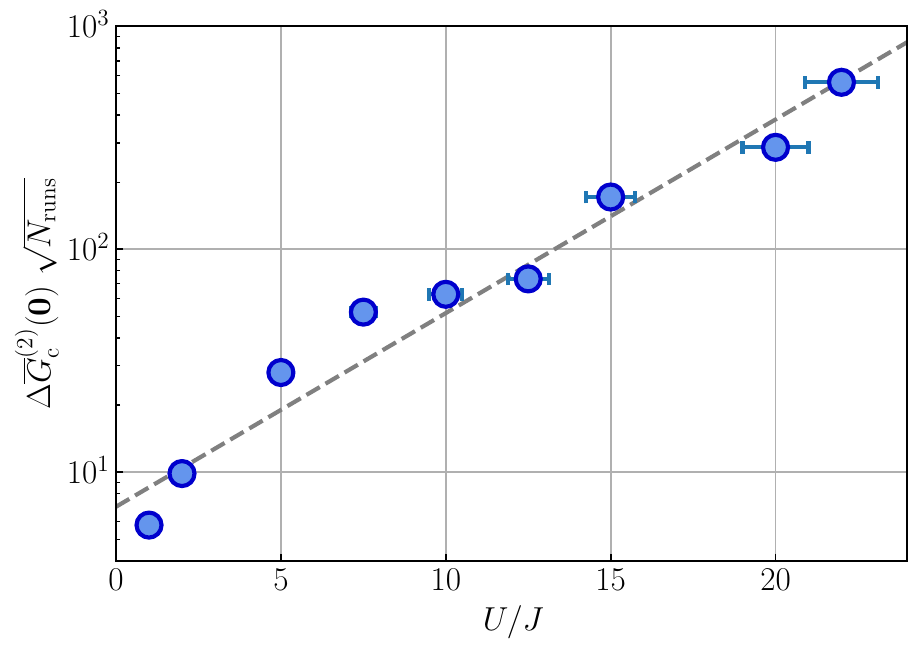}
  \caption{ Width $\Delta \overline{G}_{\rm c}^{(2)}( \bm 0)$ of the 1$\sigma$ bootstrap confidence interval on $\overline{G}_{c}^{(2)}(0)$ versus interactions $U/J$. The value of  $\Delta \overline{G}_{\rm c}^{(2)}( \bm 0)$ is multiplied by $\sqrt{N_{\rm runs}}$ where $N_{\rm runs}$ is the number of experimental runs used to compute correlations at each value of $U/J$.}. 
  \label{Fig-errorbarG2c}
\end{figure}

As it can be seen in the Fig.~2d of the main text, the error bars on the values $\overline{G}^{(2)}_{c}({\bf 0})$ strongly increase with the ratio $U/J$. In Fig.~\ref{Fig-errorbarG2c}, we plot the value of the error bars on $\overline{G}^{(2)}_{c}({\bf 0})$ normalised to the total number of experimental shots: the bare values $\Delta \overline{G}_{\rm c}^{(2)}( \bm 0)$ equal to the width of the 1$\sigma$ bootstrap confidence interval are multiplied by $\sqrt{N_{\rm runs}}$ where $N_{\rm runs}$ is the number of experimental runs used to compute the connected correlation (these vary slightly with $U/J$). We find that the variation of the error bars on $\overline{G}^{(2)}_{c}({\bf 0})$ is compatible with an exponential increase with $U/J$. 

Incidentally, we notice that the error bars on the value of two-body connected correlations contain information about higher-order of correlations since $\left( \Delta \overline{G}_c^{(2)}(0) \right)^2 = \sum_{k,k'} \left[ \langle N_k N_{-k} N_{k'} N_{-k'} \rangle - \langle N_k N_{-k} \rangle \langle N_{k'} N_{-k'} \rangle \right]$. In the future, it would be interesting to investigate whether the rapid increase reported in Fig.~\ref{Fig-errorbarG2c} reflects the contribution of these higher-order of correlations.

\section{Relation between $G^{(2)}_{c}({\bf 0})$ and four-point connected correlations $\langle a^{\dagger}_{\bf k} a^{\dagger}_{-\bf k} a_{\bf k} a_{-{\bf k}} \rangle_{c}$ }
 
We discuss the relation between the measured  two-body connected correlations $G^{(2)}_{c}({\bm k},-{\bm k})$ and four-point connected correlations $\langle a^{\dagger}_{\bm k}  a^{\dagger}_{-\bm k} a_{\bm k}  a_{-\bm k}  \rangle_{c}$ in the condensate depletion. 

The four-point connected correlation (or cumulant) $\langle a^{\dagger}_{\bm k}  a^{\dagger}_{-\bm k} a_{\bm k}  a_{-\bm k}  \rangle_{c}$ is the difference between the four-point correlations $\langle a^{\dagger}_{\bm k}  a^{\dagger}_{-\bm k} a_{\bm k}  a_{-\bm k}  \rangle$ and all connected correlations of order less or equal to $n \leq 3$. For any four operators $X_{j |j=1..4}$, the four-point connected correlations can be expressed as a function of connected correlators of order $n\leq 3$, 
\begin{eqnarray}
\langle X_{1} X_{2} X_{3} X_{4} \rangle_{c} &=& \langle X_{1} X_{2} X_{3} X_{4}  \rangle \nonumber \\
&-& \Sigma_c^{3} 
\nonumber \\
&-& \Sigma_c^{2}
\nonumber \\
&-& \langle X_{1}   \rangle_{c}  \langle X_{2} \rangle_{c}  \langle  X_{3}  \rangle_{c}  \langle X_{4}  \rangle_{c}. \nonumber
\end{eqnarray}
where the term $\Sigma_c^{3}$ containing three-body connected correlations is
\begin{eqnarray}
\Sigma_c^{3}&=& \langle X_{1}  X_{2}  X_{3}  \rangle_{c}   \langle  X_{4}  \rangle_{c}
+\langle X_{4}  X_{1}  X_{2}  \rangle_{c}   \langle  X_{3}  \rangle_{c} \\
&+&\langle X_{3}  X_{4}  X_{1}  \rangle_{c}   \langle  X_{2}  \rangle_{c}
+\langle X_{2}  X_{3}  X_{4}  \rangle_{c}   \langle  X_{1}  \rangle_{c}\nonumber \\
\end{eqnarray}
and that $\Sigma_c^{2}$ containing two-body connected correlations writes
\begin{eqnarray}
\Sigma_c^{2}&=& \langle X_{1}  X_{2}   \rangle_{c}   \langle   X_{3} X_{4}  \rangle_{c}
+\langle  X_{1}  X_{3}  \rangle_{c}   \langle  X_{2} X_{4}  \rangle_{c} \\
&+&\langle X_{1}  X_{4}    \rangle_{c}   \langle  X_{2} X_3 \rangle_{c}. \nonumber 
\end{eqnarray}
Note that $\langle X_{j}   \rangle_{c}=\langle X_{j}   \rangle$.

In the condensate depletion, the average value of one operator is null, $\langle a_{\bm k} \rangle = \langle a^{\dagger}_{\bm k} \rangle = 0 $. Inserting this fact in the above-mentioned general expression of four-point connected correlations leads to,
\begin{eqnarray}
\langle a^{\dagger}_{\bm k}  a^{\dagger}_{-\bm k} a_{\bm k}  a_{-\bm k}  \rangle_{c} = \langle a^{\dagger}_{\bm k}  a^{\dagger}_{-\bm k} a_{\bm k}  a_{-\bm k}  \rangle \nonumber \\
-  | \langle a^{\dagger}_{\bm k}  a^{\dagger}_{-\bm k}   \rangle |^2 
- | \langle  a^{\dagger}_{\bm k}  a_{-\bm k}  \rangle |^2
- \langle a^{\dagger}_{\bm k}   a_{\bm k}  \rangle  \langle  a^{\dagger}_{-\bm k}  a_{-\bm k}  \rangle. \nonumber
\end{eqnarray}

Since 
$G^{(2)}_{c}({\bm k},-{\bm k}) = \langle a^{\dagger}_{\bm k}  a^{\dagger}_{-\bm k} a_{\bm k}  a_{-\bm k}  \rangle -  \langle  a^{\dagger}_{-\bm k}  a_{-\bm k}  \rangle$
we obtain,
\begin{eqnarray}
G^{(2)}_{c}({\bm k},-{\bm k})&=& | \langle a^{\dagger}_{\bm k} a^{\dagger}_{-\bm k}  \rangle |^2 + | \langle  a^{\dagger}_{\bm k}a_{-\bm k} \rangle |^2 \nonumber \\
&+&  \langle a^{\dagger}_{\bm k}a^{\dagger}_{-\bm k} a_{\bm k} a_{-\bm k}\rangle_{c}. \label{G2c-noCoh}
\end{eqnarray}
This generic equation is valid for both Gaussian and non-Gaussian states. In the main text, we make use of this expression in the regime where Bogoliubov theory fails to show the presence of non-Gaussian correlations, {\it i.e.} $\langle a^{\dagger}_{\bm k}a^{\dagger}_{-\bm k} a_{\bm k} a_{-\bm k}\rangle_{c} \neq 0$.
\\
 
{\it Bogoliubov theory.} Within the Bogoliubov approximation which considers only linearized quantum fluctuations, quantum states are Gaussian and connected correlations of four operators are null, $\langle a^{\dagger}_{\bm k}a^{\dagger}_{-\bm k}a_{\bm k} a_{-\bm k} \rangle_{c}=0$. In addition, modes at opposite momenta are predicted to be incoherent, {\it i.e.} $| \langle  a^{\dagger}_{\bm k} a_{-\bm k} \rangle |^2 = 0$. Under these additional assumptions, the above equation Eq.~\ref{G2c-noCoh} simplifies to
\begin{equation}
G^{(2)}_{c}({\bm k},-{\bm k})_{\rm Bogo} = | \langle a^{\dagger}_{\bm k} a^{\dagger}_{-\bm k}  \rangle |^2.
\label{e.G2Bogo}
\end{equation}
This illustrates that the connected correlations $G^{(2)}_{c}({\bm k},-{\bm k})_{\rm Bogo}$ directly reflect the pairing at opposite momenta, $\langle a^{\dagger}_{\bm k} a^{\dagger}_{-\bm k}  \rangle$, under the assumptions of Bogoliubov theory.

\section{Three-body connected correlations in the experiment}

\begin{figure}[h!]
  \centering
  \includegraphics[width=\linewidth]{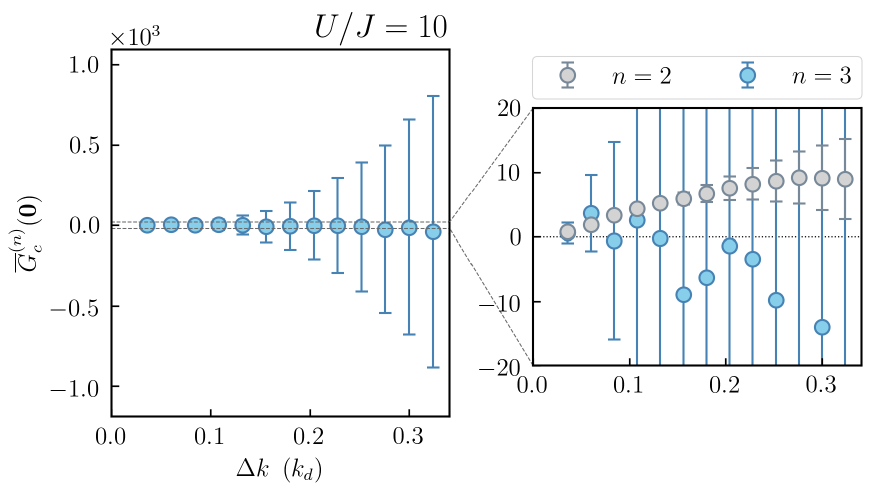}
  \caption{Three-body connected correlation $\bar{G}^{(3)}_c(\bm 0)$ as a function of the size of the integration volume $\Delta k$ for $U/J = 10$. The error bars correspond to 68\%-confidence intervals obtained from a bootstrap method. Inset: zoom into the left panel with a comparison to $\bar{G}^{(2)}_c(\bm 0)$.}
  \label{fig-sup3body}
\end{figure}

As discussed in the main text, three-body connected correlations $\overline{G}_c^{(3)}(\mathbf{0}) = \sum_{\bm k_1, \bm k_2 \in \Omega_k}  G_c^{(3)}(\bm k_1, \bm k_2, -\bm k_1 -\bm k_2)$ are ideally suited to reveal momentum-correlated triplets induced by interaction. We have computed $\overline{G}_c^{(3)}(\mathbf{0})$  from the experimental data, using the definition
\begin{align}
G_c^{(3)}(\bm k_1, \bm k_2, \bm k_3 ) & =   \langle N(\bm k_1) N(\bm k_2) N(\bm k_3) \rangle  \nonumber \\
& -  \langle N(\bm k_1) \rangle \langle N(\bm k_2) N(\bm k_3) \rangle  \nonumber \\
& -  \langle N(\bm k_2) \rangle \langle N(\bm k_1) N(\bm k_3) \rangle  \nonumber \\
& -  \langle N(\bm k_3) \rangle \langle N(\bm k_1) N(\bm k_2) \rangle  \nonumber \\
& + 2 \langle N(\bm k_1) \rangle \langle N(\bm k_2) \rangle \langle N(\bm k_3) \rangle. \label{Eq:def-G3c}
\end{align}
Unfortunately, the experimental signal-to-noise appears not sufficient to reveal the presence of a peak in $\overline{G}_c^{(3)}(\mathbf{0})/ \Delta k^3$ over the range $\Delta k < 0.2 k_d$, which would be the analog of that observed in $\overline{G}_c^{(2)}(\mathbf{0})/ \Delta k^3$ for two-body correlations (see Fig.~\ref{fig-sup3}). We illustrate this fact in figure~\ref{fig-sup3body} by plotting the measured $\overline{G}_c^{(3)}(\mathbf{0})$ at the value $U/J=10$.
 
We also compare $\overline{G}_c^{(3)}(\mathbf{0})$ with the measured values $\overline{G}_c^{(2)}(\mathbf{0})$. From our numerical calculations with quantum rotors, we expect their ratio to be maximum for $U/J=10$, with $\overline{G}_c^{(3)}(\mathbf{0})/\overline{G}_c^{(2)}(\mathbf{0})\sim 0.1$ (see fig.~\ref{fig-ratioG3G2}). Figure~\ref{fig-sup3body} illustrates that error bars on $\overline{G}_c^{(3)}(\mathbf{0})$ are too large to reveal such a signal.

\section{Correlation functions in momentum space and their scaling behavior}

\subsection{Definitions of the correlation functions}

In this work, we consider the following connected correlations of the (momentum) density operator $N(\bm k)= a^{\dagger}_{\bm k} a_{\bm k}$: $G_c^{(2)}(\bm k_1, \bm k_2) = \langle N(\bm k_1) N(\bm k_2) \rangle -  \langle N(\bm k_1) \rangle \langle N(\bm k_2) \rangle $ and $G_c^{(3)}(\bm k_1, \bm k_2, \bm k_3 )$ as defined in Eq.~\ref{Eq:def-G3c}, and their integrals over zero-sum (quasi-)momentum modes
 \begin{align} 
 \overline{G}_c^{(2)}(\mathbf{0}) = & \sum_{\bm k \in \Omega_k}  G_c^{(2)}(\bm k, - \bm k) \label{e.G2} \\ 
 \overline{G}_c^{(3)}(\mathbf{0}) = & \sum_{\bm k_1, \bm k_2 \in \Omega_k}  G_c^{(3)}(\bm k_1, \bm k_2, -\bm k_1 -\bm k_2) \label{e.G3}
 \end{align}
 
  \noindent
where the sum in Eq.~\eqref{e.G2} is defined over (quasi-) momenta $\bm k$ contained in $\Omega_k$, such that ${\bm k}+{\bm Q} \neq {-\bm k}$, where ${\bm Q}$ is a reciprocal lattice wavevector; while the sum in Eq.~\eqref{e.G3} is defined over inequivalent pairs of (quasi-) momenta $\bm k_1, \bm k_2$, namely ${\bm k}_1 \neq  {\bm k}_2 + \bm Q$, both contained in $\Omega_k$ and such that $\bm k_3 = -\bm k_1 - \bm k_2$ is also contained in $\Omega_k$.  
 
\subsection{Comparison between theory and experiment}

All theoretical calculations are performed on cubic lattices with periodic boundary conditions. In the theoretical calculations, the wavevectors ${\bm k}$ are associated with discrete {\it quasi}-momentum modes living in the first Brillouin zone, and $N({\bm k})$ are the populations of this discrete set of modes.

We define the total particle number in the region $\Omega_k$ as $\langle N_{\Omega_k} \rangle = \sum_{\bm k \in \Omega_k} \langle N(\bm k)\rangle$, and the number of modes in the region of interest as $V_{\Omega_k} = \sum_{\bm k \in \Omega_k} 1 $.
 We expect that  $\overline{G}_c^{(2)}(0)$ scales as the square of the  density of particles in $\Omega_k$, $\rho_{\Omega_k} = \langle N_{\Omega_k}\rangle /V_{\Omega_k}$, and linearly in the number of modes $\Omega_k$, namely $\overline{G}_c^{(2)}(0) \sim \langle N_{\Omega_k}\rangle ^2/V_{\Omega_k}$. Similarly we expect  $\overline{G}_c^{(3)}(0)$ to scale as the cube of the density and as 
 the square of the number of modes, so that $\overline{G}_c^{(3)}(0) \sim \langle N_{\Omega_k}\rangle ^3/V_{\Omega_k}$. As a consequence a meaningful comparison between theory and experiment referring to different particle and mode numbers requires a renormalization of the theoretical $\overline{G}_c^{(2)}(0) $ value by a factor $\sim ( \langle N_{\Omega_k}\rangle_{\rm exp}/\langle N_{\Omega_k}\rangle_{\rm th})^2 (V_{\Omega_k,\rm th}/V_{\Omega_k,\rm exp})$; and of the theoretical $\overline{G}_c^{(3)}(0) $ value by a factor 
 $\sim ( \langle N_{\Omega_k}\rangle_{\rm exp}/\langle N_{\Omega_k}\rangle_{\rm th})^3 (V_{\Omega_k,\rm th}/V_{\Omega_k,\rm exp})$. The ratio $r_V = V_{\Omega_k,\rm th}/V_{\Omega_k,\rm exp}$ is purely geometrical and independent of the Hamiltonian parameters. Yet estimating it unambiguously is not obvious. Only the core of the atomic cloud is expected to be superfluid at the temperature of the experiment and to contribute to the measured momentum-space correlations; and it is this core which in principle dictates the size of the momentum modes in momentum space. On the other hand, the ratio $r_N = \langle N_{\Omega_k}\rangle_{\rm exp}/\langle N_{\Omega_k}\rangle_{\rm th}$ depends significantly on the interaction strength and can be unambiguously determined. We have therefore decided to rescale the theoretical data for $\overline{G}_c^{(2)}(0)$ and $\overline{G}_c^{(3)}(0)$ by the factors  $r_N^2$ and  $r_N^3$, respectively, assuming that $r_V \sim O(1)$. Indeed we observe that the rescaled theoretical curve $r_N^2 \overline{G}_c^{(2)}(0)$ matches the experimental one in order of magnitude.
 
 \begin{figure}[h!]
  \centering
  \includegraphics[width=\linewidth]{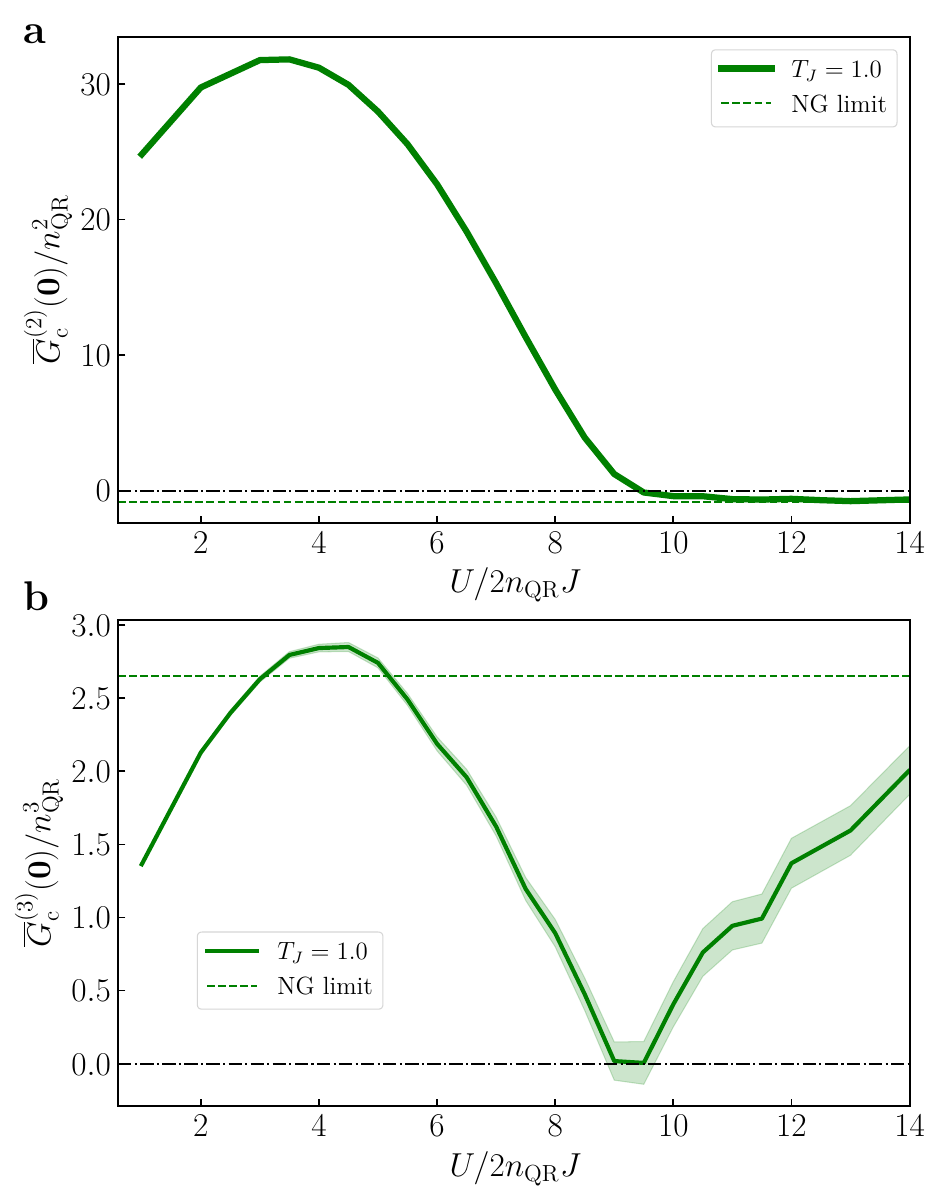}
  \caption{
  {\bf a} $\overline{G}^{(2)}(0)$ integrated correlations for the quantum-rotor model with size $V=10^3$ for the $\Omega_k$ region defined as in the main text, and for a temperature $T_{J}=T/(2Jn_{\rm QR}) = 1.0$. The solid line indicates the deep normal-gas limit. {\bf b}   $\overline{G}^{(3)}(0)$ integrated correlations for the quantum-rotor model with size $V=6^3$ for an $\Omega_k$ region corresponding to a cubic corona contained within two cubes with sides $k =2 k_{d} \pi/3$ and $k = k_{d}$, and at a temperature $T_{J}=T/(2Jn_{\rm QR}) = 1.0$. The solid line indicates again the deep normal-gas limit.}
  \label{fig:G2G3}
\end{figure}

 \subsection{Scaling of the correlation functions in the superfluid vs. in the Mott (or normal gas) limit}

We discuss here the different scaling regimes of the momentum correlations predicted by the theory on quantum rotors. We restrict our attention to a system in the {\it canonical} ensemble at fixed particle number $N$, and at fixed lattice filling $n_{\rm QR} = N/V$ where $V=L^3$ is the number of sites. 
 In the superfluid regime we expect the two-mode correlations $G_c^{(2)}(\bm k, - \bm k)$ to be $\sim O(1)$, as predicted by Bogoliubov theory, and therefore we expect $\overline{G}_c^{(2)}(0) \sim O(N)$ since $V_{\Omega_k} \sim V$. This scaling is indeed observed in our quantum-rotor calculations. On the other hand we do not know of a simple theoretical prediction for $ G_c^{(3)}(\bm k_1, \bm k_2, -\bm k_1 -\bm k_2)$ -- indeed Bogoliubov theory predicts this type of correlation to be vanishing. Hence we resort to numerics in order to estimate the scaling of $\overline{G}_c^{(3)}(0) $, and we find from our quantum-rotor calculations that $\overline{G}_c^{(3)}(0) \sim O(N)$. This implies that three-mode correlations are much more tenuous than two-mode ones, namely   $G_c^{(3)}(\bm k_1, \bm k_2, \bm k_3) \sim O(N^{-1})$ so that their integral over $O(N^2)$ trios of modes is $O(N)$.  
 
The scaling of the integrated correlations in the superfluid regime should be contrasted with that of a deep Mott-insulating or normal-gas regime, in which all quasi-momentum modes become statistically independent, modulo the conservation of the particle number. The population of each mode is thermally distributed so as to exhibit bunching, with average occupation $\langle N(\bm k)\rangle  = n_{\rm QR} $. 
Moreover one has $\langle N^2(\bm k)\rangle = n_{\rm QR} (2n_{\rm QR}+1)$ and  $ \langle N^3(\bm k)\rangle = 6n_{\rm QR}^3 + 6n_{\rm QR}^2 + n_{\rm QR}$, according to thermal statistics. A lengthy but straightforward calculation leads to the result
 \begin{align}
 G_c^{(2)}(\bm k_1, \bm k_2) & \to - \frac{n_{\rm QR}(n_{\rm QR}+1)}{V-1} \nonumber \\
  G_c^{(3)}(\bm k_1, \bm k_2, \bm k_3)  & \to \frac{4 n_{\rm QR}^3 + 6  n_{\rm QR}^2 +  2n_{\rm QR}}{(V-1)(V-2)}
 \end{align} 
(for inequivalent quasi-momenta  $\bm k_1$, $\bm k_2$ and $\bm k_3$). This implies that, for  $n_{\rm QR} \gg 1$ (see Sec.~\ref{s.QR})
 \begin{align}
 \overline{G}_c^{(2)}(0)  & \to   - \frac{ n_{\rm QR}^2 ~V_{\Omega_k}}{V-1}  \nonumber \\
 \overline{G}_c^{(3)}(0)  & \to   \frac{4  n_{\rm QR}^3 ~{\cal N}_{3,\Omega_k}}{(V-1)(V-2)}
 \end{align}
 where ${\cal N}_{3,\Omega_k}$ is the number of momentum trios $\bm k_1$, $\bm k_2$ and $-\bm k_1-\bm k_2$ contained in $\Omega_k$. Since $V_{\Omega_k} \sim V$ and ${\cal N}_{3,\Omega_k} \sim V^2$, both quantities take limiting values which do {\it not} scale with the particle number, marking a strong suppression of correlations with respect to the superfluid regime.
 
 Fig.~\ref{fig:G2G3} shows results for the $\overline{G}_c^{(2)}(0)$ and  $\overline{G}_c^{(3)}(0)$ correlations of the quantum-rotor model  without any further renormalization factor.
 We observe that the asymptotic behavior predicted in the deep normal-gas limit at large interaction strength is correctly reproduced by our quantum rotor results. As explained in the next section, this is to be expected \cite{carcy2019}, since in the deep normal-gas limit the momentum distribution of the quantum rotor model becomes a speckle pattern, whose exponential statistics \cite{Dainty-book} is the same as the one of thermal Bose fields. Interestingly, the integrated  $\overline{G}_c^{(3)}(0)$  does not reach this limit monotonically from above, but it rather shows an intermediate minimum, corresponding roughly to the superfluid/normal-gas transition. 

\subsection{Ratio of three-body and two-body connected correlations}

\begin{figure}[ht!]
  \centering
  \includegraphics[width=\linewidth]{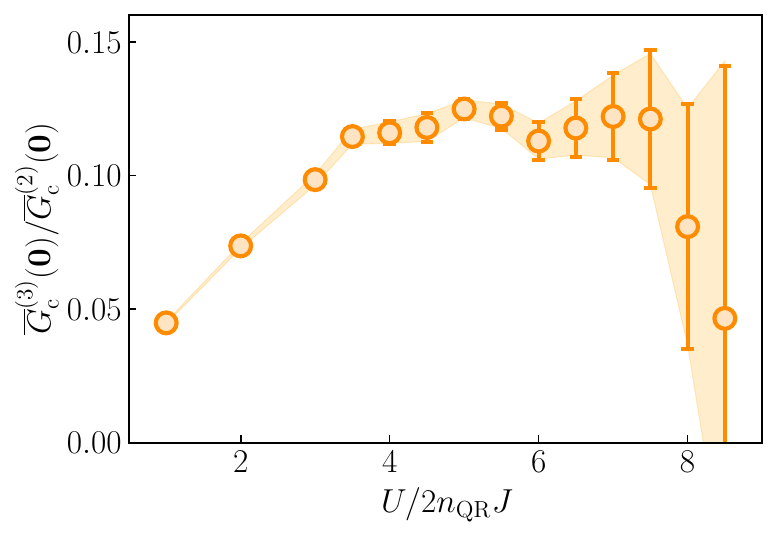}
  \caption{Ratio $G_{c}^{(3)}(0)/G_{c}^{(2)}(0)$ versus $U/2n_{\rm QR}J$ obtained with the model of quantum rotors.
  \label{fig-ratioG3G2}
  }
\end{figure}

We discuss here the variation of the amplitudes of two-body and three-body connected correlations obtained with the quantum rotor results. In Fig.~\ref{fig-ratioG3G2}, we plot the ratio $G_{c}^{(3)}(0)/G_{c}^{(2)}(0)$ as a function of $U/2n_{\rm QR}J$. We identify two regimes, $U/2n_{\rm QR}J<4$ and $U/2n_{\rm QR}J\geq 4$. 
 
At small values of interactions $U/2n_{\rm QR}J<4$, both the amplitudes $G_{c}^{(3)}(0)$ and $G_{c}^{(2)}(0)$ increase with interaction (see main text). However, the increase of their ratio illustrates that $G_{c}^{(3)}(0)$ increases faster than $G_{c}^{(2)}(0)$. This finding is compatible with the microscopic picture depicted in Fig.~1 of the main text: as the condensate fraction decreases, the importance of the interaction term $V^{(3)}$ responsible for the emergence of correlated triplets grows.

At larger interactions $U/2n_{\rm QR}J\geq 4$, we find that the ratio $G_{c}^{(3)}(0)/G_{c}^{(2)}(0)$ is roughly constant. This indicates that the perturbative picture drawn in Fig.~1 of the main text does not apply to the equilibrium state at large interactions. This is compatible with the fact that, in this regime, the condensate fraction is small and that the various terms depicted in Fig. 1 have similar contributions.
 
\subsection{Behavior of the $G_c^{(2)}$ correlations at the superfluid/normal-gas transition}

As seen in the previous subsection, $G_c^{(2)}(0)$ turns to negative values in the deep normal-gas limit, while it is positive in the deep superfluid regime because of Bogolyubov momentum pairing. How does $G_c^{(2)}(0)$ transition between these two regimes? 

\begin{figure}[h!]
  \centering
  \includegraphics[width=\linewidth]{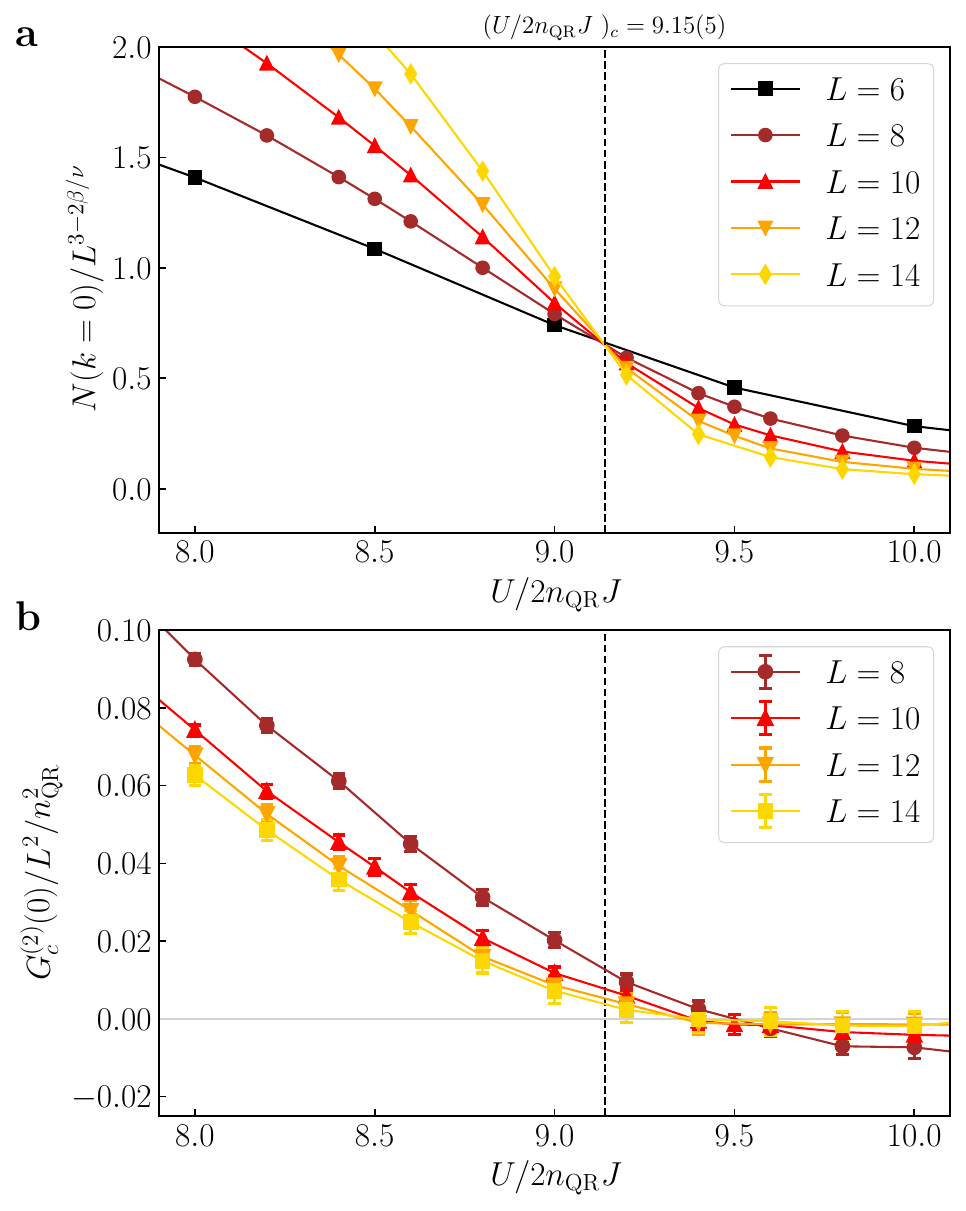}
  \caption{Superfluid/normal-gas transition for the quantum-rotor model at $T/(2Jn_{\rm QR}) = 1$. Upper panel: scaling behavior of the condensate fraction at the SF/NG transition, identifying the critical point at $[U/(2Jn_{\rm QR})]_c$ Here $\beta = 0.3485$ and $\nu = 0.67155$ are critical exponents of the 3d XY universality class. Lower panel: behavior of $G_c^{(2)}(0)$ in the vicinity of the transition for $L^3$ cubic lattices. $G_c^{(2)}(0)$ is defined by integrating over a cubic corona with momentum modes $(2\pi/L)(n_x,n_y,n_z)$ with $|n_\alpha| \geq L/2-2$ for a least one component $\alpha = x, y$ or $z$.}
  \label{fig-sup6}
\end{figure}

Fig.~\ref{fig-sup6} shows quantum-rotor results for the critical behavior of $G_c^{(2)}(0)$ at the SF/NG transition driven by an increasing interaction at fixed temperature. The position of the transition is identified via the scaling of the condensate fraction. When examining the behavior of $G_c^{(2)}(0)$ in the vicinity of the transition one observes that it remains positive over the entire SF regime, and that it appears to vanish in the vicinity of the transition, to then turn negative in the NG phase. In particular the value of $U/(2Jn_{\rm QR})$ at which $G_c^{(2)}(0)$ crosses zero appears to move closer to the critical value as the system size increases, suggesting that the critical point may correspond to the point at which $G_c^{(2)}(0)$ changes sign. 

\section{Finite-temperature Bogoliubov theory}

The results from Bogoliubov theory presented in the main text have been obtained by a simple self-consistent generalization of the theory to access finite temperature.  Bogoliubov theory is based on the following approximate grand-canonical Hamiltonian on a 3D cubic lattice
\begin{eqnarray}
H - \mu N & \approx & \frac{UN_0^2}{2V} - \mu N_0 + \sum_{\bm k \neq 0} (e_{\bm k} + 2Un_0 -\mu ) a^\dagger_{\bm k} a_{\bm k} \nonumber \\
& + & \frac{Un_0}{2} \sum_{\bm k \neq 0} \left ( a_{\bm k} a_{-\bm k} + {\rm h.c.} \right )   
\end{eqnarray}
where $e_{\bm k} = -2J \left [ \cos(k_x) + \cos(k_y) + \cos(k_z) \right ]$, and $n_0 = N_0/V$ is the condensate density. 
The chemical potential $\mu$ is fixed by requiring the condensate to satisfy the Gross-Pitaevskii equation, which leads to the condition $\mu = Un _0$. 

The Bogoliubov Hamiltonian is then diagonalized by the transformation $b_{\bm k} = u_{\bm k} a_{\bm k} + v_{\bm k} a^\dagger_{-\bm k}$ to the form 
\begin{equation}
H - \mu N \approx \sum_{\bm k\neq 0} \epsilon_{\bm k} b_{\bm k}^\dagger b_{\bm k} + {\rm const.}
\end{equation}
where $\epsilon_{\bm k} = \sqrt{ A_{\bm k}^2 - B_{\bm k}^2} = \sqrt{ e_{\bm k} (  e_{\bm k} + 2Un_0 )}$ with $A_{\bm k} = e_{\bm k} + Un_0$, $B_{\bm k} = Un_0$, $u_{\bm k} = \sqrt{\frac{1}{2} (A_{\bm k}/E_{\bm k} + 1)}$, and $v_{\bm k} =  \sqrt{\frac{1}{2} (A_{\bm k}/E_{\bm k} - 1)}$. 
The population of finite-momentum modes at finite temperature reads then
\begin{equation}
\langle a_{\bm k}^\dagger a_{\bm k} \rangle = v_{\bm k}^2 +  n_B(\epsilon_{\bm k}, T) (u_{\bm k}^2 + v_{\bm k}^2) 
\end{equation}
where $n_B(\epsilon_{\bm k}, T) = (e^{\epsilon_{\bm k}/(k_B T)} -1 )^{-1}$ is the Bose distribution for Bogoliubov quasiparticles. This expression depends parametrically on the condensate density $n_0$ which enters into the dispersion relation $\epsilon_{\bm k}$ as well as in the $u_{\bm k}$ and $v_{\bm k}$ coefficients. 
Fixing the total density at temperature $T$
\begin{equation}
n = n_0 + \frac{1}{V} \sum_{\bm k\neq 0} \langle a_{\bm k}^\dagger a_{\bm k} \rangle
\label{e.density}
\end{equation}
amounts then to imposing a non-linear equation on the condensate density $n_0$ (or equivalently on the chemical potential), which can be solved numerically. In the example given in the main text we have chosen $n=1$. 

Once the condensate density satisfying Eq.~\eqref{e.density} is found, one can evaluate the pairing correlation 
\begin{equation}
\langle a^{\dagger}_{\bm k} a^{\dagger}_{-\bm k}  \rangle = -u_{\bm k} v_{\bm k} \left [ 1 + 2n_B(\epsilon_{\bm k}, T) \right ]
\end{equation}
leading to the connected correlations of Eq.~\eqref{e.G2Bogo}, shown in the main text.

\section{Quantum-rotor model and Quantum Monte Carlo} \label{s.QR}

The quantum-rotor model is an approximation to the Bose-Hubbard model valid at large, integer filling $n_{\rm QR}$ \cite{Wallinetal1994}. Its starting point is the amplitude-phase decomposition $a_i = e^{i\phi} \sqrt{n_i}$ of the bosonic field, with $[\phi, n_i] = 1$.  Rewriting the density as $n_i = n_{\rm QR} + \delta n_i$, and assuming that  $\sqrt{\langle (\delta n_i)^2 \rangle} \ll n_{\rm QR} $ (since $n_{\rm QR} \gg 1$) in all physical regimes of interest, then one can write 
\begin{equation}
a_i^\dagger a_j + {\rm h.c.} \approx 2 n_{\rm QR} \cos(\phi_i - \phi_j)
\label{e.approx}
\end{equation}
neglecting density fluctuations. Density-density interactions are nonetheless introducing quantum fluctuations of the phase, since 
\begin{equation}
n_i^2 = -\frac{\partial^2}{\partial \phi_i^2}  ~.
\end{equation}
Working at integer filling eliminates all linear terms in the density from the Hamiltonian, leading to the quantum rotor Hamiltonian
\begin{equation}
H_{\rm QR} = -2Jn_{\rm QR} \sum_{\langle ij \rangle}  \cos(\phi_i - \phi_j) - \frac{U}{2} \sum_i \frac{\partial^2}{\partial \phi_i^2}~.
\end{equation}
Using a coherent-state path-integral approach \cite{Wallinetal1994, SondhiG1997}, the quantum statistical physics of the quantum-rotor Hamiltonian in $d$ dimensions can be mapped onto an effective classical problem of rotors i.e. planar spins, or angles $\phi_{i,p}$, where $i$ is the spatial position and $p$ is the position in the extra (Trotter) dimension. A quantum Monte Carlo simulation of the quantum rotor model amounts to sampling the configurations of this $(d+1)$-dimensional classical model \cite{Wallinetal1994, Roscildeetal2016}. Given the approximation of Eq.~\eqref{e.approx}, to each configuration of the classical rotors $\{\phi_{i,p} \}$ at a given Trotter step $p$ one can uniquely associate a momentum distribution 
\begin{equation}
N_p(\bm k) = \frac{n_{\rm QR}}{V} \sum_{ij} e^{i \bm k \cdot \left ( {\bm r}_i - {\bm r}_j \right )} e^{-i \left ( \phi_{i,p} - \phi_{j,p} \right )} ~.
\end{equation}
Hence sampling the partition function of the classical rotors allows for the reconstruction of arbitrary correlation functions of the momentum populations $N(\bm k)$. This aspect is at the core of our use of the quantum rotor model to probe the fluctuations of the momentum distribution. 

In particular, in the deep normal-gas limits the quantum-rotor variables $\phi_i$ become uncorrelated random angles, so that the quantum-rotor picture for the Bose field in momentum space
\begin{equation}
b_{\bm k} \to \sqrt{\frac{n_{\rm QR}}{V}} \sum_i e^{i\bm q \cdot \bm r_i} e^{i\phi}
= \frac{1}{\sqrt{V}} \sum_{i} \sqrt{n_{\rm QR}} ~e^{i\theta_i} 
\end{equation}
becomes a sum of random complex numbers of fixed amplitude $\sqrt{n_{\rm QR}}$ and random phase $\theta_i = \bm q \cdot \bm r_i + \phi_i$. The momentum distribution is then the square modulus of this random complex variable, which, in the limit of a large system $V \gg 1$, has the same statistics of the intensity of a laser-speckle pattern \cite{Dainty-book}, namely 
\begin{equation}
P(N_{\bm k}) \to  P (|b_{\bm k}|^2) = \frac{e^{- \frac{N_{\bm k}}{n_{\rm QR}}}}{n_{\rm QR}}~,    
\end{equation}
namely the same statistics as that of thermal photons in the limit of a large average photon number, $n_{\rm QR}\gg 1$.

The original Bose-Hubbard model of interest can also be studied in its equilibrium thermodynamics via different quantum Monte Carlo approaches \cite{trotzky2010, SSE}. Yet all these approaches use Fock states in real space as computational basis, and therefore the momentum distribution is an off-diagonal observable. Estimating off-diagonal observables requires a dedicated update scheme (a so-called "worm algorithm"). Complex worm-like algorithms have been developed in the past by some of us to probe the second moment of the fluctuations of the momentum distribution in the one-dimensional Bose-Hubbard model \cite{Roscilde2008,Fangetal2016}, but third moments have never been probed to the best of our knowledge. An important direction for the future will then be to develop such dedicated algorithms for the study of the fluctuations of the momentum distribution of the three-dimensional Hubbard model.

\end{document}